\documentclass[12pt,a4wide,epsf]{article}
\usepackage{epsf}
\usepackage[hang,nooneline]{subfigure}

\usepackage{graphicx}
\newcommand{\insertplot}[5]{\begin{figure}
 \hfill\hbox to 0.05in{\vbox to #5in{\vfill
 \inputplot{#1}{#4}{#5}}\hfill}
 \hfill\vspace{-.1in}
 \caption{#2}\label{#3}
 \end{figure}}
 \newcommand{\inputplot}[3]{
 \special{ps: plotfile #1}
}
\newcounter{fig}   
\begin{document}

\title{
Hairy Wormholes
and
Bartnik-McKinnon Solutions
}

\vspace{1.5truecm}
\author{
{\bf Olga Hauser$^1$},
{\bf Rustam Ibadov$^2$},
{\bf Burkhard Kleihaus$^1$},
{\bf Jutta Kunz$^1$}\\
$^1$
Institut f\"ur  Physik, Universit\"at Oldenburg\\ Postfach 2503,
D-26111 Oldenburg, Germany\\
$^2$
Department of Theoretical Physics and Computer Science,\\
Samarkand State University, Samarkand, Uzbekistan
}

\vspace{1.5truecm}

\date{\today}

\maketitle
\vspace{1.0truecm}

\begin{abstract}
We consider Lorentzian wormholes supported by a phantom field
and threaded by non-trivial Yang-Mills fields,
which may be regarded as hair on the Ellis wormhole.
Like the Bartnik-McKinnon solutions and their associated
hairy black holes, these hairy wormholes form infinite sequences,
labeled by the node number $k$ of their gauge field function.
We discuss the throat geometry of these wormholes,
showing that odd-$k$ solutions may exhibit a double-throat,
and evaluate their global charges.
We analyze the limiting behavior exhibited by wormhole solutions
as the gravitational coupling becomes large.
The even-$k$ solutions
approach smoothly the Bartnik-McKinnon solutions with $k/2$ nodes,
while the odd-$k$ solutions develop a singular behavior at the throat
in the limit of large coupling.
In the limit of large $k$, on the other hand,  
an embedded Abelian wormhole solution is approached,
when the throat is large.
For smaller throats the extremal Reissner-Nordstr\"om solution
plays a fundamental role in the limit.
\end{abstract}


\vfill\eject

\section{Introduction}

The discovery of a sequence of spherically symmetric solutions 
of SU(2) Einstein-Yang-Mills theory
by Bartnik and McKinnon \cite{Bartnik:1988am}
came as a surprise,
since the very existence of globally regular solutions
was unexpected. 
These Bartnik-McKinnon solutions are asymptotically flat,
and they are characterized by
the node number $k$ of their purely magnetic gauge field.

To every regular spherically symmetric Bartnik-McKinnon solution
there exists a corresponding family of black hole solutions
with regular event horizon 
\cite{Volkov:1990sva,Bizon:1990sr,Kuenzle:1990is}.
Since these black hole solutions possess 
non-trivial non-Abelian fields outside their event horizon,
they represent counterexamples to the ``no-hair'' conjecture.
They are hairy black holes carrying Yang-Mills hair.

Instead of black holes we here consider wormholes and
address the question, whether we can endow these wormholes
in a similar way with Yang-Mills fields, and thus with hair.
The counterpart of the Schwarzschild solution,
which corresponds to the hairless
solution in the case of the hairy black holes,
will now be the Ellis wormhole
\cite{Ellis:1973yv,Ellis:1979bh,Bronnikov:1973fh},
while the hairy wormholes will have non-trivial
non-Abelian fields threading their throat.

The Ellis wormhole is based on the presence of a phantom field,
i.e., a scalar field with
a reversed sign in front of its kinetic term
\cite{Ellis:1973yv,Ellis:1979bh,Bronnikov:1973fh,Kodama:1978dw,ArmendarizPicon:2002km}.
Such a phantom field provides the required violation of the
energy conditions, necessary to obtain traversable wormholes
in Einstein gravity
\cite{Morris:1988cz}.
Nowadays phantom fields are ubiquitous in cosmology,
since they could explain the observed accelerated expansion
of the Universe
\cite{Lobo:2005us}.

Wormholes with phantom fields have been studied in numerous
respects (see e.g.~\cite{Visser}),
including the presence of wormholes inside stars and neutron stars
\cite{Dzhunushaliev:2011xx,Dzhunushaliev:2012ke,Dzhunushaliev:2013lna},
or the study of their shadow
\cite{Bambi:2013nla,Nedkova:2013msa}.
However, wormholes can also be obtained without phantom fields
by allowing higher curvature terms in the action
\cite{Hochberg:1990is,Fukutaka:1989zb,Ghoroku:1992tz,Furey:2004rq,Bronnikov:2009az,Kanti:2011jz,Kanti:2011yv} 
or by coupling matter non-minimally to Einstein gravity
\cite{Balakin:2007xq,Balakin:2010ar}.

In fact, wormholes with non-minimally coupled Yang-Mills fields
were constructed in \cite{Balakin:2007xq}.
But these solutions possess only a simple Wu-Yang type gauge field.
Thus they correspond to wormhole solutions carrying
magnetic charge. The possibility of finding sequences of wormholes 
threaded by genuine non-Abelian fields was not explored, however.
Instead, black wormholes were introduced, whose throat
is hidden behind an event horizon.

The wormholes we consider here also allow for charged
Wu-Yang type solutions. These correspond to embedded Abelian
solutions, where the gauge field function is identically zero. 
When the gauge field is pure gauge, 
on the other hand, 
the Ellis wormhole is recovered.
Our interest focusses, however, on wormholes with non-trivial
gauge field functions, which can be labeled by a node number $k$,
and we consider only symmetric wormholes.

In particular,
we analyze the behavior of odd-$k$ and even-$k$ wormholes,
and demonstrate that
the even-$k$ solutions approach smoothly the Bartnik-McKinnon solutions,
when the gravitational coupling tends to infinity.
In contrast,
the odd-$k$ solutions exhibit a singular behavior in this limit
in the vicinity of the throat, while outside the throat region
they also approach Bartnik-McKinnon solutions (for $k>1$).
We also discuss the limit $k \to \infty$,
where the analogy with hairy black holes becomes particularly
apparent.

The paper is organized as follows.
In section 2 we discuss the action, the Ans\"atze and the equations of motion.
We analyze the wormhole properties
and discuss the violations of the energy conditions.
Moreover, we briefly recall the exact Ellis wormhole 
and discuss the exact charged Abelian wormhole.
In section 3 we present our numerical results. 
We first discuss the probe limit, and then present the gravitating solutions.
In particular, we discuss the solutions themselves, 
their throat geometry, their global charges, and their dependence
on the coupling constant.
Subsequently, we discuss the limiting behavior of the wormholes
for large coupling and for large node number.
We end with our conclusions in section 4.

\section{The model}

\subsection{Action and field equations}

We consider Einstein gravity coupled to a phantom field
and an SU(2) Yang-Mills field. The action
\begin{equation}
S=\int \left[ \frac{1}{16\pi G}R 
+  L_{\rm ph} 
+  L_{\rm YM} \right] \sqrt{-g} d^4x
\ \label{action}  \end{equation}
consists of the Einstein-Hilbert action with curvature scalar $R$,
Newton's constant $G$, and the determinant of the metric $g$,
and of the respective matter contributions.
There are the Lagrangian of the phantom field $\phi$
\begin{equation}
 L_{\rm ph} =  \frac{1}{2} \partial_\mu \phi \partial^\mu \phi
\ , \end{equation}
and the Yang-Mills Lagrangian
\begin{equation}
L_{\rm YM}= -\frac{1}{2} {\rm Tr} (F_{\mu\nu} F^{\mu\nu})
\ , \end{equation}
with field strength tensor $F_{\mu\nu}$
\begin{equation}
F_{\mu\nu}= \partial_\mu A_\nu - \partial_\nu A_\mu
            - i   [A_\mu,A_\nu]
\ , \end{equation}
gauge potential $A_\mu$
\begin{equation}
A_\mu = \frac{1}{2} \tau^a A_\mu^a
\ , \end{equation}
and Pauli matrices $\tau^a$.
Note, that the gauge coupling constant is set to one.

Variation of the action with respect to the metric
leads to the Einstein equations
\begin{equation}
G_{\mu\nu}= {\cal R}_{\mu\nu}-\frac{1}{2}g_{\mu\nu}{\cal R} 
= \bar \alpha T_{\mu\nu}
\label{ee} 
\end{equation}
with gravitational coupling constant $\bar \alpha = 8\pi G$ and
stress-energy tensor
\begin{equation}
T_{\mu\nu} = g_{\mu\nu}{{\cal L}}_M
-2 \frac{\partial {{\cal L}}_M}{\partial g^{\mu\nu}} \ ,
\label{tmunu} 
\end{equation}
where ${\cal L}_{\rm M} = 
{\cal L}_{\rm ph}+{\cal L}_{\rm YM}$ is the matter Lagrangian.
 
Variation with respect to the matter fields leads to
the gauge field equations
\begin{equation}
\frac{1}{\sqrt{-g}} D_\mu(\sqrt{-g} F^{\mu\nu}) = 0 \ ,
\label{feqA} 
\end{equation}
where $D_\mu = \partial_\mu - i [ A_\mu, \cdot \, ]$,
and the phantom field equation
\begin{equation}
\partial_\mu \left( \sqrt{-g} \,\partial^\mu \phi\right) =0 \ .
\label{feqp}
\end{equation}

\subsection{Ans\"atze}

To construct  static spherically symmetric wormhole solutions
we employ the line element
\begin{equation}
ds^2 = -A^2 dt^2 +d\eta^2 + f N d\Omega^2 \ ,
\label{lineel}
\end{equation}
where $d\Omega^2=d\theta^2 +\sin^2\theta d\varphi^2$
denotes the metric of the unit sphere, while $A$ and $N$ are functions of
$\eta$, and 
\begin{equation}
f=\eta^2 + \eta_0^2 
\end{equation}
is an auxiliary function. 
We note, that the coordinate $\eta$ takes positive and negative
values, i.e.~$-\infty< \eta < \infty$. 
The limits $\eta\to \pm\infty$
correspond to two distinct asymptotically flat regions.

The Ansatz for the SU(2)-gauge potential is chosen 
as for the globally regular Bartnik-McKinnon solutions.
The gauge potential has vanishing time component,
$A_0=0$, and the spatial components involve the Pauli matrices
and a single profile function $K(\eta)$
\begin{eqnarray}
A_i &=& \frac{1-K(\eta)}{2\eta  } (\vec e_\eta \times \vec \tau)_i
\ . \label{su2} \end{eqnarray}
The phantom field $\phi$ depends only on the coordinate $\eta$, as well.

Substitution of these Ans\"atze into the Einstein equations leads to
\begin{eqnarray}
0 & = & 
\bar \alpha \frac{f^2 N^2 \phi'^2 - 2 N f K'^2 - (K^2-1)^2}{N^2 f^2} 
+\frac{\eta^2}{f^2}
\nonumber\\ & & 
 +\frac{4 N (1 - 3 \eta N' -2 N) - f(4 N N'' - N'^2)}{4 f N^2} \ ,
\label{eineq00} \\
0 & = & 
\bar  \alpha \frac{f^2 N^2\phi'^2 - 2 f N K'^2 + (K^2-1)^2}{N^2 f^2}
 + \frac{\eta^2}{f^2}
\nonumber\\ & & 
+ \frac{f(A N'^2+ 4N A' N') + 4(\eta A  N N' - A N+2 \eta A' N^2)
       }{4 A f N^2} \  ,
\label{eineqrr} \\
0 & = & 
-\bar \alpha \frac{ f^2 N^2 \phi'^2+(K^2 - 1)^2}{f N} 
 - \frac{ \eta^2 N}{f}
\nonumber\\ & & 
+ \frac{f\left( 2 A  N N'' - A N'^2+ 2 A' N N'+ 4 A'' N^2 \right)}{4 A N} 
\nonumber\\ & & 
   +\frac{A N + \eta A N' + \eta A' N }{A} \  .
\label{eineqtt} 
\end{eqnarray}
for the $tt$, $\eta\eta$ and $\theta\theta$ components, respectively,
and the prime denotes differentiation with respect to $\eta$.
Substitution into the matter equations yields
\begin{equation}
K'' =\frac{A K (K^2 - 1) - A'N f K'}{A f N} \ ,
\end{equation}
\begin{equation}
\phi''=-\phi'\frac{A f N'+2 A N \eta +N f A'}{A N f}
=-\phi'\frac{(A N f)'}{A N f} \ .
\end{equation}
The last equation can be integrated
\begin{equation}
\phi' = \frac{D}{A N f} \ ,
\label{phip}
\end{equation}
where the constant $D$ is a measure for the scalar charge
of the phantom field.
Consequently, we can eliminate
the phantom field from the Einstein equations,
employing the substitution $\phi'^2 = \left(D/A f N\right)^2$.

Let us now introduce dimensionless quantities
\begin{equation}
x=\frac{\eta}{\eta_0} \ , \ \ \ 
\hat{f}=\frac{f}{\eta_0^2} = x^2+1\ , \ \ \ 
{\alpha} = \frac{\bar \alpha}{\eta_0^2} \ .
\end{equation}
By subtracting Eq.~(\ref{eineqrr}) from Eq.~(\ref{eineq00})
and from Eq.~(\ref{eineqtt}) we obtain 
our final set of equations, which has the form
\begin{eqnarray}
.N'' & = & 
-2{\alpha}\frac{(K^2-1)^2}{\hat{f}^2 N} 
+  \frac{2 A(1- N -2 x N') - A'( \hat{f}N' + 2 x N)}{A \hat{f}} \ ,
\label{eqN}\\
A''& = & 
{\alpha} A \frac{2N\hat{f}K'^2 + (K^2-1)^2}{\hat{f}^2 N^2} 
-A'\frac{\hat{f} N' + 2x N}{\hat{f}N} \ ,
\label{eqA}\\
K'' & = &  \frac{A K(K^2-1) - N A'\hat{f} K'}{AN\hat{f}} \ ,
\label{eqK}\\
D^2  & = & 
A^2 (2N\hat{f} K'^2 - (K^2-1)^2) 
\nonumber\\ & & 
+ \frac{A}{4{\alpha}}
\left[
\hat{f}^2 ( -A N'^2- 4 N A' N') 
+4\hat{f} (AN - 2 x A' N^2 - x A N N') \right.
\nonumber\\ & & 
\left. \ \ \ \ \ \ \ \ 
 - 4 x^2 A N^2 \right] \ ,
\label{eqd2}
\end{eqnarray}
where the prime now denotes the derivative with respect to 
the dimensionless coordinate $x$.
We employ the condition $D=const$ to monitor the quality of the
numerical solutions.

\subsection{Wormhole throats}

In the following we consider only wormholes with a metric
that is symmetric under $x \to -x$, 
and with symmetric or antisymmetric matter field functions.
To discuss the wormhole geometry we introduce the shape function
of the wormhole $R(x)$, where
\begin{equation}
R^2(x) = N(x) f(x) = N(x) (x^2 + 1) \ .
\label{R}
\end{equation}
$R$ may be considered as a circumferential radial coordinate.
Asymptotically, the shape function should tend to the
modulus of $x$ to guarantee asymptotic flatness of the wormholes.
On the other hand, the shape function should not possess a zero
to yield a wormhole solution.
Therefore $R(x)$ should possess at least one minimum, where
it is positive and where
its derivative vanishes. 

Because of the assumed symmetry of the wormholes,
$x=0$ should be an extremum, i.e.,~$R'(0)=0$, 
and we may choose $R(0)=1$, i.e., 
\begin{equation}
N(0)=1 \ .
\label{N0}
\end{equation}
If $R(x)$ has a minimum at $x=0$, then $x=0$ corresponds to the throat
of the wormhole.
If on the other hand $R(x)$ has a local maximum at $x=0$, 
then $x=0$ corresponds to an equator.
In that case, the wormhole will have a double-throat
surrounding a belly. 
Such double-throat solutions were found before
in wormholes threaded by chiral fields \cite{Charalampidis:2013ixa}.

To find out, whether we are dealing with a single throat or a double-throat, 
we consider the condition for a minimum of the function $R$
\begin{equation}
\label{R_min}
\left. \frac{d^2 R^2}{d x^2} \right|_{x=0}
  =\left. N''f+4N'x + 2 N \right|_{x=0}
=N''(0) +2 > 0 \ ,
\end{equation}
using Eq.~(\ref{N0}). Evaluating now Eq.~(\ref{eqN}) at $x=0$
and inserting this expression for $N''(0)$ into Eq.~(\ref{R_min}), we obtain
\begin{equation}
 - 2  \alpha { ( K^2(0) - 1)^2 } + 2 > 0 \ .
\end{equation}
Thus there is a critical value of $ \alpha$, 
\begin{equation}
  \alpha_{\rm cr} = \frac{1}{ ( K^2(0) - 1)^2 } \ ,
\label{acrit}
\end{equation}
such that for $ \alpha <  \alpha_{\rm cr}$
wormholes have a single throat,
whereas for $ \alpha >  \alpha_{\rm cr}$
wormholes have a double-throat.
For wormholes with an odd number of nodes, the gauge field
function must vanish at $x=0$, i.e., for all odd $k$
the critical value is $ \alpha_{\rm cr} =1$.

\subsection{Wormhole geometry}

The shape of the wormhole can be visualized with the help of
embedding diagrams. To that end we consider the metric
at fixed $t$ and $\theta=\pi/2$
and embed it isometrically in Euclidean space
\begin{equation}
\label{embed}
d s^2 = d x^2 +fN d \phi ^2
= d \rho^2 + \rho^2 d\varphi^2 + d z^2  \ ,
\end{equation}
where $\rho = \rho(x)$, $z=z(x)$.
Comparison yields
\begin{equation}
\frac{d\rho}{d x}^2 + \frac{dz}{dx}^2 =1 \ , \quad \rho^2=f N \ ,
\end{equation}
which allows to evaluate $z(x)$,
\begin{equation}
z(x) = \int_0^x \sqrt{ 1 - \frac{1}{4} \frac{ ( ( fN )' )^2 } {fN} } dx' \ .
\end{equation}

\subsection{Boundary conditions}

To obtain globally regular solutions, 
which are asymptotically flat and possess a finite mass,
we must make an appropriate choice for the six free boundary conditions.
There are three singular points, where boundary conditions may be imposed.
These are the position of the center of the wormhole $x=0$,
and the asymptotic infinities $x = \pm \infty$.
For the metric functions we choose the boundary conditions
\begin{equation}
A'(0) = 0 \ , \quad A(\infty) = 1 \ , \quad N(0) = 1 \ .
\label{bound1}
\end{equation}
We may also choose $N'(0)=0$ instead of $A'(0) = 0$.
The remaining conditions for asymptotic flatness, 
$A(-\infty) = 1$ and $N(\pm \infty)=1$,
are satisfied automatically.

For the gauge field function we impose asymptotically
the same boundary conditions
as for the Bartnik-McKinnon solutions, i.e.
\begin{eqnarray}
&&K(-\infty) = \pm 1 \ , \quad K(+\infty) = \mp 1 \ \ {\rm for \ odd} \ k \ ,
\nonumber \\
&&K(-\infty) = \pm 1 \ , \quad K(+\infty) = \pm 1 \ \ {\rm for \ even} \ k \
\label{bound2}
\end{eqnarray}
and supplement these with a condition at the center of the wormhole
\begin{equation}
K(0) = 0 \ \ {\rm for \ odd} \ k \ , \quad K'(0) = 0  \ \ {\rm for \ even} \ k \ .
\label{bound3}
\end{equation}

\subsection{Energy conditions}

Since violation of the null energy condition (NEC)
implies violation of the weak and strong energy conditions,
we here focus on the NEC, which states that
\begin{equation}
\Xi = T_{\mu\nu} k^\mu k^\nu \ge 0 \ ,
\end{equation}
for all (future-pointing) null vector fields $k^\mu$.

Employing the Einstein equations, 
this condition can be expressed 
for spherically symmetric solutions as
\begin{equation}
-G_0^0+G_l^l \geq  0 \ , \ \ \ {\rm and } 
\ \ \ -G_0^0+G_\theta^\theta\geq  0 \ .
\label{Nulleng}
\end{equation}
If one or both of the above conditions 
do not hold in some region of spacetime, then the
null energy condition is violated.
This is the case for all solution studied.

\subsection{Wormhole properties}

The area of the throat ${\cal A}_{\rm th}$ is determined
by
\begin{equation}
{\cal A}_{\rm th} = 4 \pi R^2(\eta_{\rm th}) \ .
\end{equation}
For a single throat wormhole the throat is located at
$\eta_{\rm th}=0$. The area of the throat is then given by
\begin{equation}
{\cal A}_{\rm th} = 4 \pi \eta^2_0 = 4 \pi \frac{\bar \alpha}{\alpha} \ .
\end{equation}
Thus for fixed $\bar \alpha$ and
increasing $ \alpha$ the throat shrinks in size.
For a double-throat wormhole, on the other hand, $\eta=0$ corresponds to the
location of a maximal surface, while the locations 
of the two throats $\pm \eta_{\rm th}$ 
depend on the node number and the coupling constant $ \alpha$.

The surface gravity $\kappa$ at the throat is determined via
\begin{equation}
\kappa^2 = -1/2 \left. (\nabla_\mu \xi_\nu)(\nabla^\mu \xi^\nu) \right|_{\eta_{\rm th}} \ ,
\end{equation}
where $\xi^\mu$ is the timelike Killing vector field,
i.e., for the metric Ansatz (\ref{lineel}) 
\begin{equation}
\kappa =  A'(\eta_{\rm th}) \ .
\label{kap}
\end{equation}
Symmetric wormholes with a single throat have vanishing $\kappa$,
for double-throat wormholes $\kappa$  is finite, however.

The mass of the wormhole solutions can be read off the metric function
$A$ at infinity,
\begin{equation}
A^2 \to 1 - \frac{2 G M}{\eta} = 1 - \frac{2 \mu}{x} \ ,
\label{mass}
\end{equation}
i.e., the mass parameter $\mu$ is given by
\begin{equation}
\mu =  \alpha \eta_0 \frac{M}{8 \pi} \ .
\end{equation}
Alternatively, one can use the Komar integral to obtain the mass
\begin{equation}
M = M_{\rm th} +\frac{1}{4\pi G} \int_\Sigma{ R_{\mu\nu} \xi^\mu n^\nu} dV
  = M_{\rm th} -\frac{1}{4\pi G} \int{ R^0_{0}\sqrt{-g}}d^3x \ , 
\label{M_adm}
\end{equation}
where $\Sigma$ is a spacelike hypersurface ($\eta_{\rm th} \le \eta \le \infty$),
$n^\nu$ is a normal vector on $\Sigma$,
$dV$ is the natural volume element on $\Sigma$,
and $M_{\rm th}$ denotes the contribution of the throat to the mass
\begin{equation}
M_{\rm th} = \frac{ \kappa A_{\rm th}}{4 \pi G} \ .
\label{m_th}
\end{equation}
$M_{\rm th}$ contributes only for double-throat wormholes.

The scalar charge $D$ determines the asymptotic behavior of the
phantom field and is obtained from Eq.~(\ref{eqd2}).
A gauge invariant definition of the non-Abelian magnetic charge
is given in Ref.~\cite{Corichi:1999nw,Corichi:2000dm}
\begin{equation}
 {\cal P}^{\rm YM} = \frac{1}{4\pi}
 \oint \sqrt{\sum_i{\left(F^i_{\theta\varphi}\right)^2}} d\theta d\varphi
 =  |P|
\ , \label{Pdelta} \end{equation}
where the integral is evaluated at spatial infinity,
yielding $P=0$ for the hairy wormholes.

\subsection{Exact solutions}\label{exact}

The model has exact wormhole solutions.
The first exact solution corresponds to the Ellis wormhole.
Here the metric functions are constant, the gauge field vanishes,
and the phantom field is an elementary function
\begin{equation}
A(x) = N(x) = 1 \ , \quad 
K(x) = \pm 1 \ , \quad
\phi(x) = D \arctan (x) + \phi_0 \ ,
\end{equation}
where we choose the integration constant $\phi_0=-D \pi/2$.
This solution represents the analogue of the Schwarzschild solution
for black holes, since no gauge field is present.
Clearly, the Ellis wormhole is massless.

The second exact wormhole solution carries charge. Here the gauge field
corresponds to the Wu-Yang monopole, i.e., 
the gauge potential is given by Eq.~(\ref{su2}) with $K(x)=0$.
This Wu-Yang gauge field solves the gauge field equation trivially,
and represents an embedded U(1) solution with unit magnetic charge.
The black hole analogon of this solution corresponds to
the embedded Reissner-Nordstr\"om solution with unit charge.

The U(1) charged wormholes were presented in \cite{Gonzalez:2009hn}.
Based on the line element 
\begin{equation}
ds^2 = -e^{2 d}dt^2 + 
e^{-2 d}\left[ d\bar{r}^2 + (\bar{r}^2+b^2) d\Omega^2\right] \ ,
\label{metricmex}
\end{equation}
the symmetric wormhole solutions are given by \cite{Gonzalez:2009hn}
\begin{equation}
e^d = \frac{\cos\left(\frac{\lambda}{b}\frac{\pi}{2}\right)}{\cos(\lambda \xi)} \ , \ \ \ 
\phi = D\left( \xi-\frac{\pi}{2b}\right) \ , 
\label{solmex}
\end{equation}
where $\xi = \frac{1}{b}\arctan(\bar{r}/b)$, 
and the constant $\lambda$ and $b$ obey the condition
\begin{equation}
\lambda^2 - \bar \alpha \cos^2\left(\frac{\lambda}{b}\frac{\pi}{2}\right)\ = 0 \ .
\label{condlam}
\end{equation}
Thus the range of $\lambda$ is restricted to $0 \leq \lambda < 1$.

The areal radius of the throat is
$R = b e^{-d(0)} = b /\cos\left(\frac{\lambda}{b}\frac{\pi}{2}\right)$.
To compare with Yang-Mills wormhole solutions with $R=\eta_0$
the constants should obey the condition
\begin{equation}
b-\eta_0\cos\left(\frac{\lambda}{b}\frac{\pi}{2}\right)= 0 \ . 
\label{condb}
\end{equation}
Together with Eq.~(\ref{condlam}) this yields
\begin{equation}
b = 
  \eta_0 \cos\left(\sqrt{{\alpha}}\frac{\pi}{2}\right) \ , \ \ \ 
\lambda = 
  \eta_0 \sqrt{{\alpha}}\cos\left(\sqrt{{\alpha}}\frac{\pi}{2}\right)
 \ .
\label{bandlam}
\end{equation}
We observe, that in the limit $ {\alpha} \to 1$ both $b$ and $\lambda$
tend to zero, while the ratio $\lambda/b$ tends to one.

\begin{figure}[h!]
\begin{center}
\mbox{\hspace*{-1.0cm}
\subfigure[][]{
\includegraphics[height=.27\textheight, angle =0]{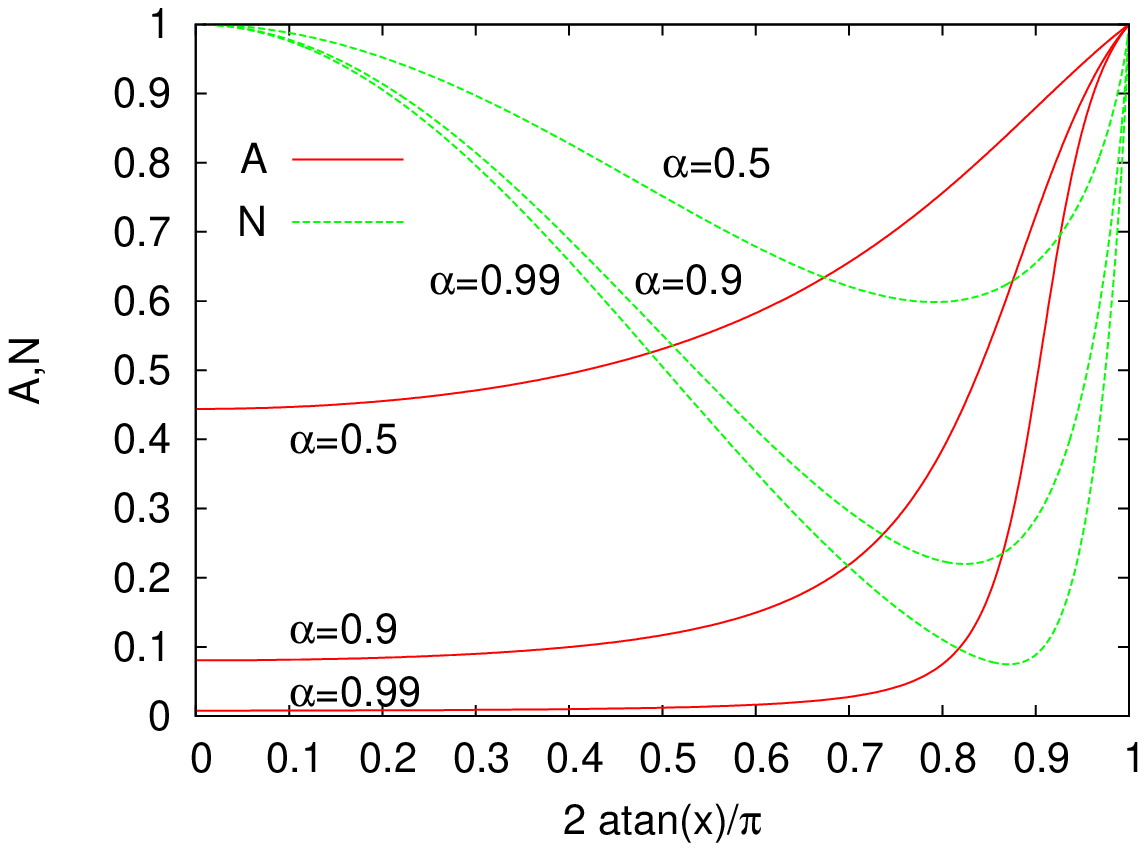}
\label{fig-1a}
}
\hspace*{-1.0cm}
\subfigure[][]{
\includegraphics[height=.27\textheight, angle =0]{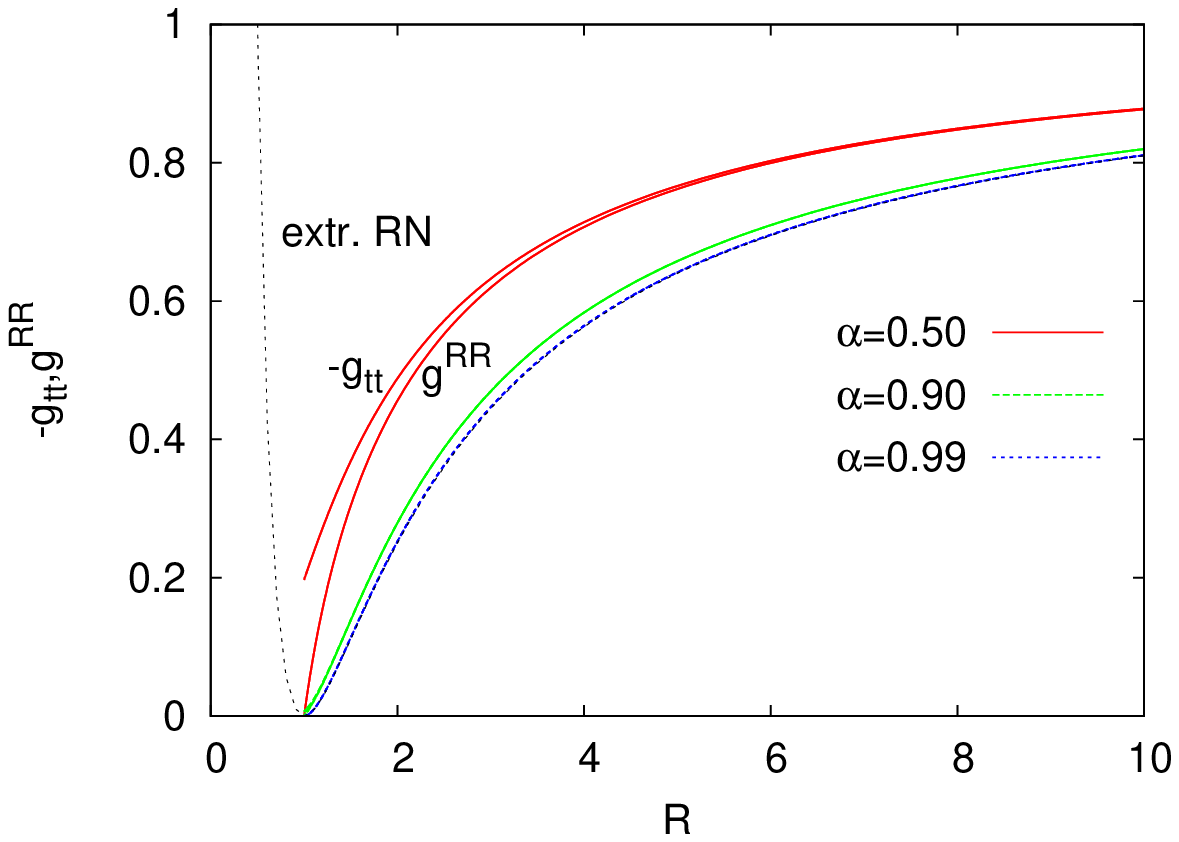}
\label{fig-1b}
}
}
\end{center}
\vspace*{-0.5cm}
\caption{\small
The wormhole metric functions $A$ and $N$ 
are shown versus the radial coordinate $x$
(a),  
and the metric functions $g^{RR}$ and $g_{tt}$ are exhibited versus 
the Schwarzschild-like coordinate $R$
(b) for $ \alpha=0.5$, 0.9 and 0.99.
The latter are compared to the metric functions of the
extremal Reissner-Nordstr\"om solution.
}
\label{fig-1}
\end{figure}

For later comparison with the non-Abelian wormhole solutions
we also evaluate the metric functions $A$ and $N$ of these Abelian wormhole
solutions and exhibit them in Fig.~\ref{fig-1a}
versus the radial coordinate $x$ for several values of the
coupling parameter $ \alpha$.
The range of existence of these Abelian wormhole solutions
is $0  <  \alpha < 1$.
When $ \alpha \to 1$ the solutions become singular in these coordinates.

To identify the limiting solution obtained
for $ \alpha \to 1$ let us transform to the
Schwarzschild-like coordinate $R$
with line element 
\begin{equation}
ds^2 = g_{tt} dt^2 + g_{RR} dR^2 + R^2 d\Omega^2 \ .
\end{equation}
Comparison with Eq.~(\ref{lineel}) yields
\begin{equation}
g_{tt} = -A^2 \ , \ \ \ \ 
g^{RR} = 
\left[\frac{2\eta N +(\eta^2+\eta_0^2)N'}{2\sqrt{(\eta^2+\eta_0^2) N}} \right]^2
\ , \ \ \ \ 
R      = \sqrt{(\eta^2+\eta_0^2) N} \ ,
\label{wh_in_Sch}
\end{equation}
and the extremal Reissner-Nordstr\"om is given by
\begin{equation}
-g_{tt} = g^{RR}= \left(1-\frac{R_{\rm H}}{R}\right)^2 \ .
\label{xBH_in_Sch}
\end{equation}

The metric functions $g^{RR}$ and $g_{tt}$ 
of the Abelian wormhole solutions
are exhibited versus $R$
in Fig.~\ref{fig-1b} for the same set of values of $ \alpha$.
Here we see that both functions approach the same limiting
function, when $ \alpha \to 1$.
This limiting function is the metric function
of the extremal Reissner-Nordstr\"om solution with unit charge
outside the event horizon,
also shown in the figure.
Thus the family of Abelian wormhole solutions ends in an
extremal black hole solution.

Considering the global charges of these wormhole solutions, one obtains
the dimensionless mass $\mu$ from
the asymptotic expansion of the function $e^d$.
With condition (\ref{condlam}) this yields for the scaled dimensionless
mass
\begin{equation}
\frac{\mu}{\sqrt{{\alpha}}}=\sin(\sqrt{ \alpha} \pi/2) \ .
\end{equation}
This implies $0 \leq \frac{\mu}{\sqrt{{\alpha}}} < 1$.
The scaled scalar charge is given by
\begin{equation}
 {\alpha} D^2 
 =  \left(1-{\alpha}\right) 
              \cos^2\left(\sqrt{{\alpha}}\frac{\pi}{2}\right) \ ,
\end{equation}
and the magnetic charge by
\begin{equation}
P=1 \ . 
\end{equation}

\section{Results}

Let us now discuss the results of the numerical integration of the
system of ordinary differential equations, Eqs.~(\ref{eqN})-(\ref{eqK}),
subject to the boundary conditions (\ref{bound1})-(\ref{bound3}).
After a brief discussion of the numerical technique
we present wormhole solutions for node numbers
$k=1,...,6$ and discuss their properties.
Subsequently, we consider the limit $ \alpha \to \infty$
for odd-$k$ and even-$k$ solutions,
as well as the limit $k \to \infty$ for fixed $ \alpha$.

\subsection{Numerical method}

In the numerical calculations we take units such that
$\bar \alpha = 8 \pi G=1$. 
and employ a collocation method for boundary-value ordinary
differential equations, equipped with an adaptive mesh selection procedure
\cite{COLSYS}.
Typical mesh sizes include $10^2-10^3$ points.
The solutions have a relative accuracy of $10^{-6}$.
 
\subsection{Probe limit}

\begin{figure}[t!]
\begin{center}
\includegraphics[height=.35\textheight, angle =0]{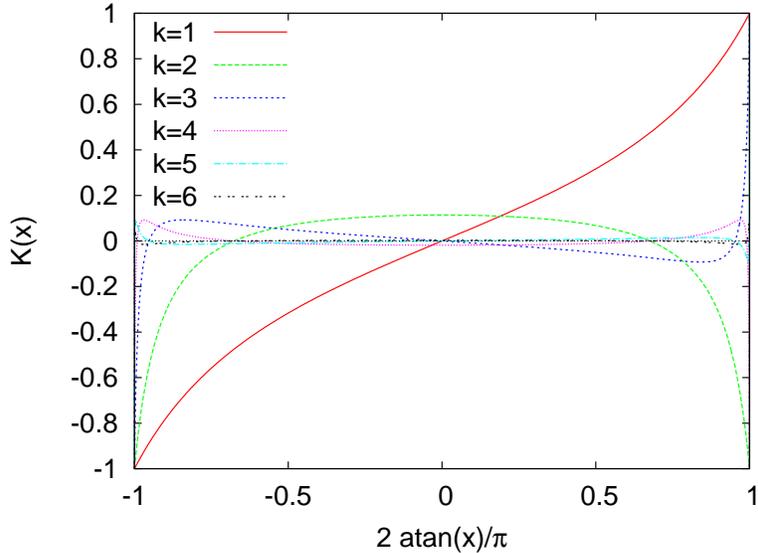}
\end{center}
\vspace*{-0.5cm}
\caption{\small
Gauge field function $K$ for node numbers $k=1,...,6$
and throat size parameter $\eta_0=1$
in the probe limit $ \alpha = 0$.
}
\label{fig0}
\end{figure}

Let us begin our discussion with the probe limit of the wormhole solutions,
obtained in the limit of vanishing $ \alpha$.
In this case the Yang-Mills field does not
contribute in the Einstein equations, and the metric
becomes the metric of the Ellis wormhole, i.e. $A=1$, $N=1$,
and the value of $\eta_0$ determines the radius of the throat.

The Yang-Mills equation then simplifies to
\begin{equation}
K'' = \frac{K(K^2-1)}{\eta^2+\eta_0^2} \ .
\label{eqKplim}
\end{equation}
In contrast to the flat space
limit, where $\eta_0=0$,
this ordinary differential equation (ODE) possesses nontrivial solutions.

The Yang-Mills equation (\ref{eqKplim}) is solved numerically in the background of
the Ellis wormhole. For any value of $\eta_0>0$,
a sequence of globally regular solutions is found, labelled by
the node number $k$ of the gauge field function $K$.
The gauge field function $K$ for the solutions with $k=1,...,6$ nodes 
and throat size $\eta_0=1$ is exhibited in Fig.~\ref{fig0}.

\subsection{A sequence of non-Abelian wormhole solutions}

Let us now take the backreaction of the gauge field on the
metric into account and consider the full system of 
coupled Einstein-Yang-Mills-scalar equations.
We recall that for a vanishing gauge potential, i.e., $K(x)=1$, 
the Ellis wormhole is obtained,
whereas for the constant gauge field function $K(x)=0$
the gauge potential corresponds to the Wu-Yang monopole, 
and an embedded Abelian wormhole with unit charge is obtained.

\begin{figure}[h!]
\begin{center}
\mbox{\hspace*{-1.0cm}
\subfigure[][]{
\includegraphics[height=.27\textheight, angle =0]{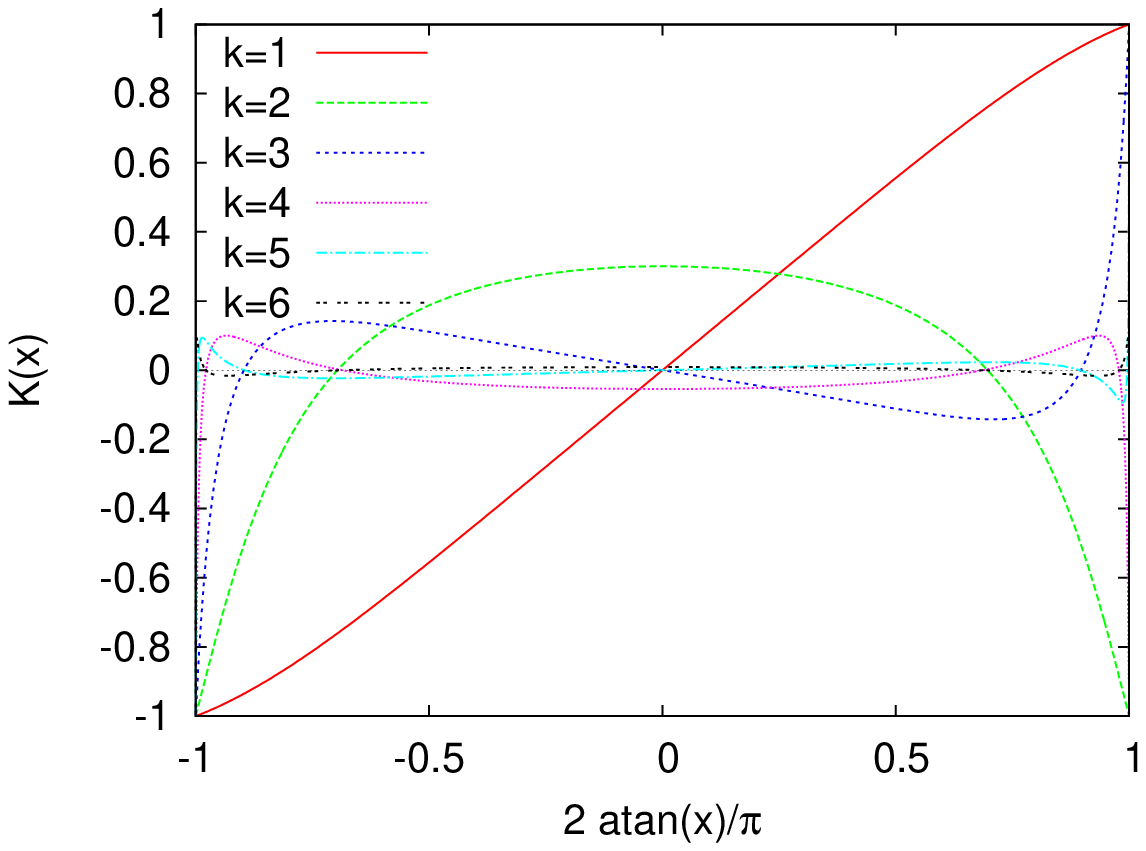}
\label{fig1a}
}
\hspace*{-1.0cm}
\subfigure[][]{
\includegraphics[height=.27\textheight, angle =0]{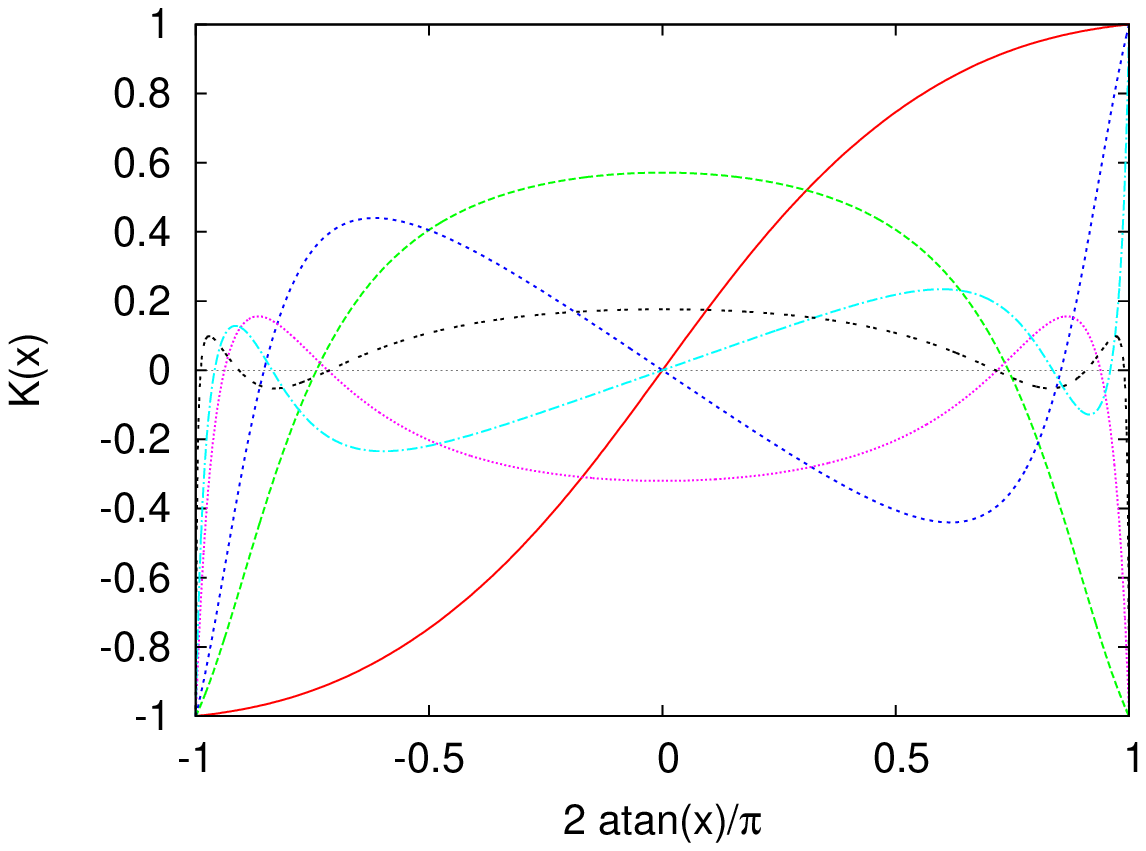}
\label{fig1b}
}
}
\mbox{\hspace*{-1.0cm}
\subfigure[][]{
\includegraphics[height=.27\textheight, angle =0]{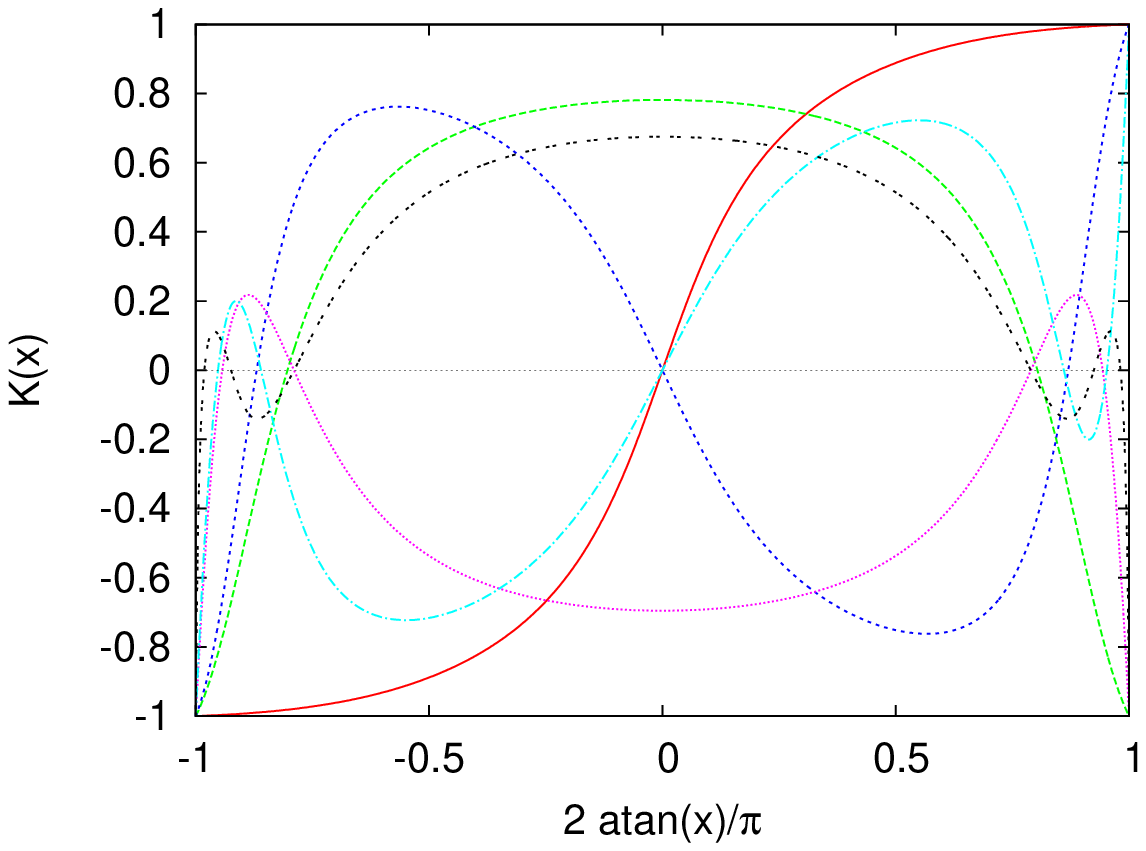}
\label{fig1c}
}
\hspace*{-1.0cm}
\subfigure[][]{
\includegraphics[height=.27\textheight, angle =0]{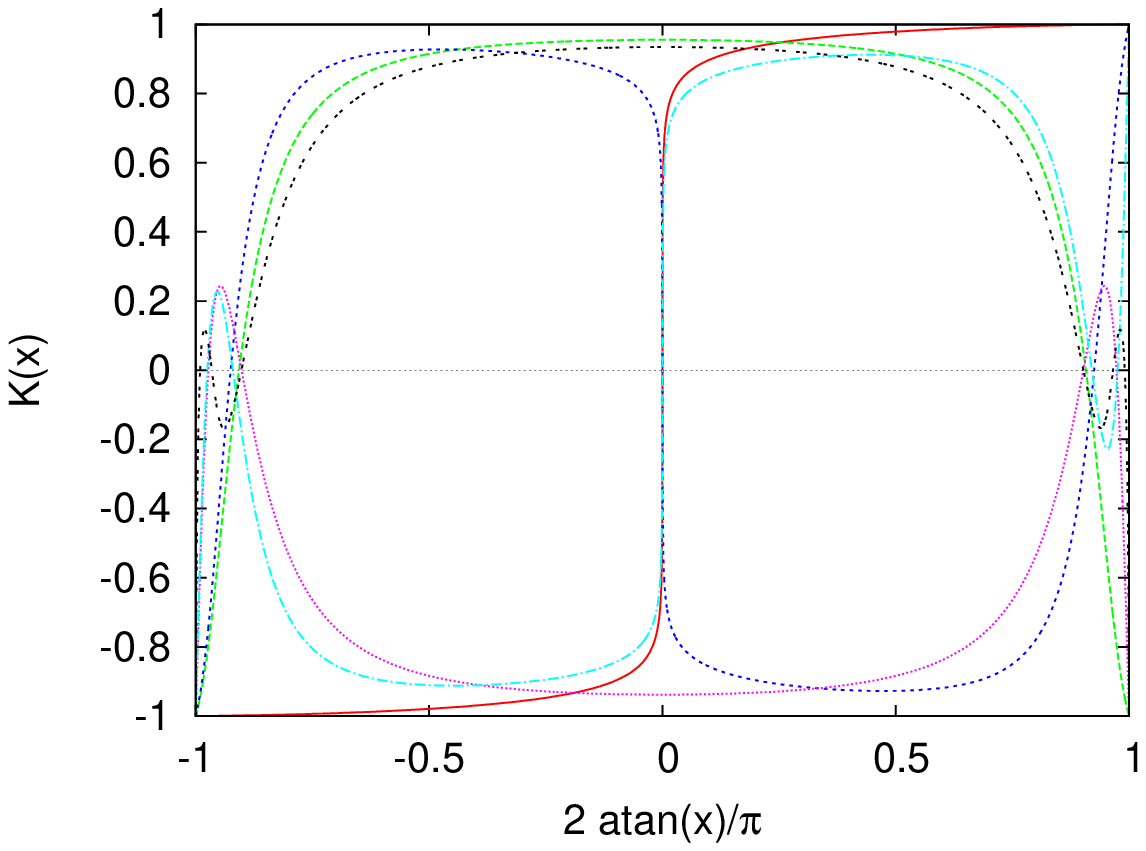}
\label{fig1d}
}
}
\end{center}
\vspace*{-0.5cm}
\caption{\small
The wormhole gauge field function $K(x)$ is shown for solutions
with $k$ nodes, $k=1,...,6$,
and $ \alpha=0.5$ (a), 1 (b), 2 (c) and 10 (d).}
\label{fig1}
\end{figure}

As we increase the coupling constant $ \alpha$ from zero,
for each node number $k$ a family of wormhole solutions
emerges smoothly from the corresponding solution
found in the probe limit.
This is seen in
Fig.~\ref{fig1}, which exhibits the gauge field function $K(x)$ 
for non-Abelian wormhole solutions with $k$ nodes, $k=1,...,6$, 
obtained for $ \alpha=0.5$, 1, 2 and 10.

The odd-$k$ solutions have $K(0)=0$ and $K'(0) \ne 0$.
For $ \alpha=1$
they correspond precisely to the solutions at the critical value
$\alpha_{\rm cr}$, where the throat is degenerate in the sense,
that for smaller values of $ \alpha$ there is a single throat at $x=0$, 
whereas for larger $ \alpha$ there are two symmetric throats
at $\pm x_{\rm th}$ with an equator inbetween at $x=0$.
The even-$k$ wormhole solutions have $K(0) \ne 0$ and $K'(0)=0$.
Taking into account Eq.~(\ref{acrit}), we note that the even-$k$ solutions 
have always $ \alpha < \alpha_{\rm cr}$
and thus possess always only a single throat.

\begin{figure}[h!]
\begin{center}
\mbox{\hspace*{-1.0cm}
\subfigure[][]{
\includegraphics[height=.27\textheight, angle =0]{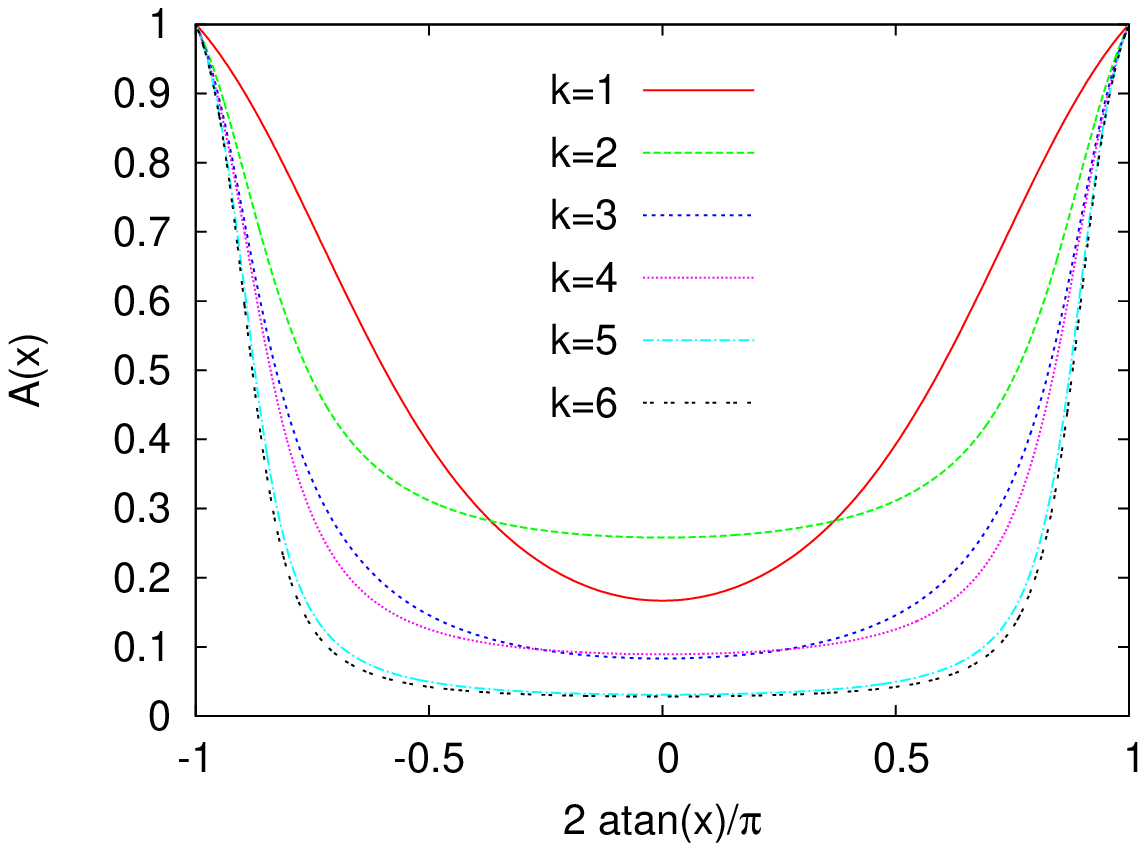}
\label{fig2a}
}
\hspace*{-1.0cm}
\subfigure[][]{
\includegraphics[height=.27\textheight, angle =0]{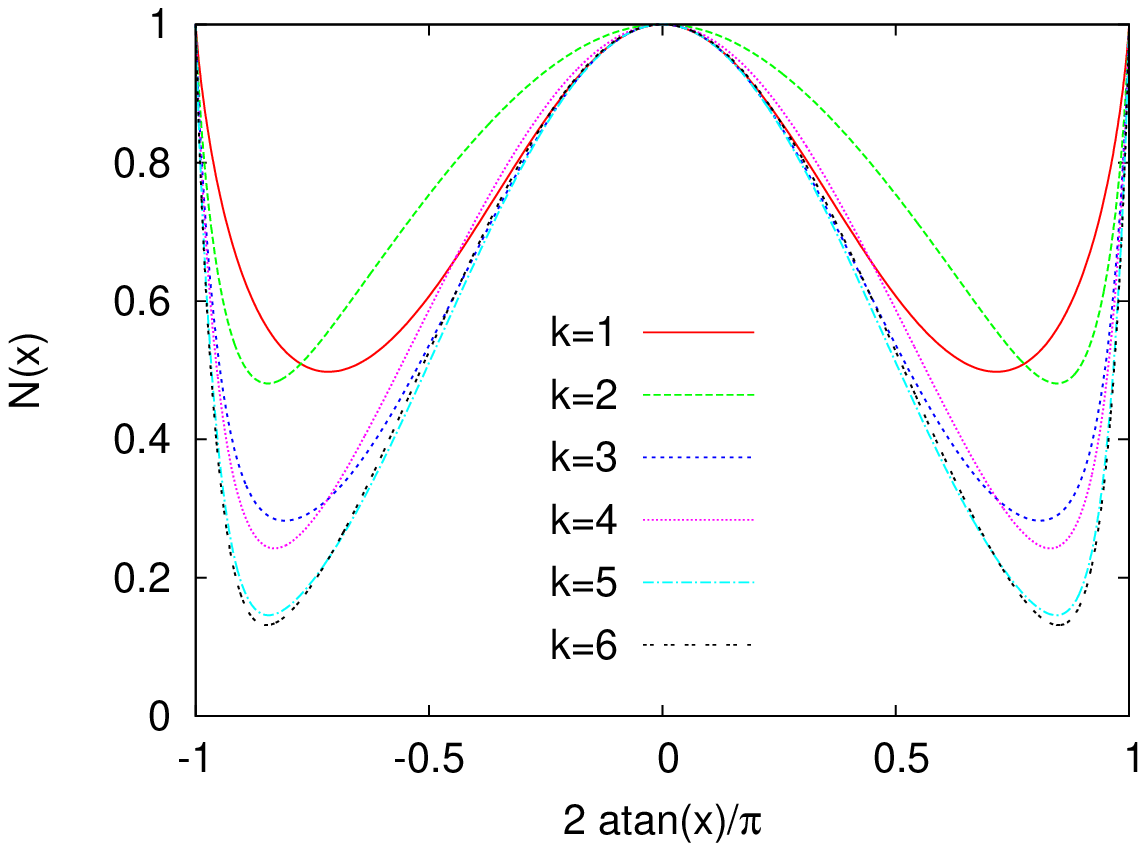}
\label{fig2b}
}
}
\end{center}
\vspace*{-0.5cm}
\caption{\small
The wormhole metric functions $A(x)$ (a) and $N(x)$ (b) are shown
for solutions with $k$ nodes, $k=1,...,6$,
and $ \alpha=1$.
}
\label{fig2}
\end{figure}

\begin{figure}[h!]
\begin{center}
\mbox{\hspace*{-1.7cm}
\subfigure[][]{
\includegraphics[height=.33\textheight, angle =0]{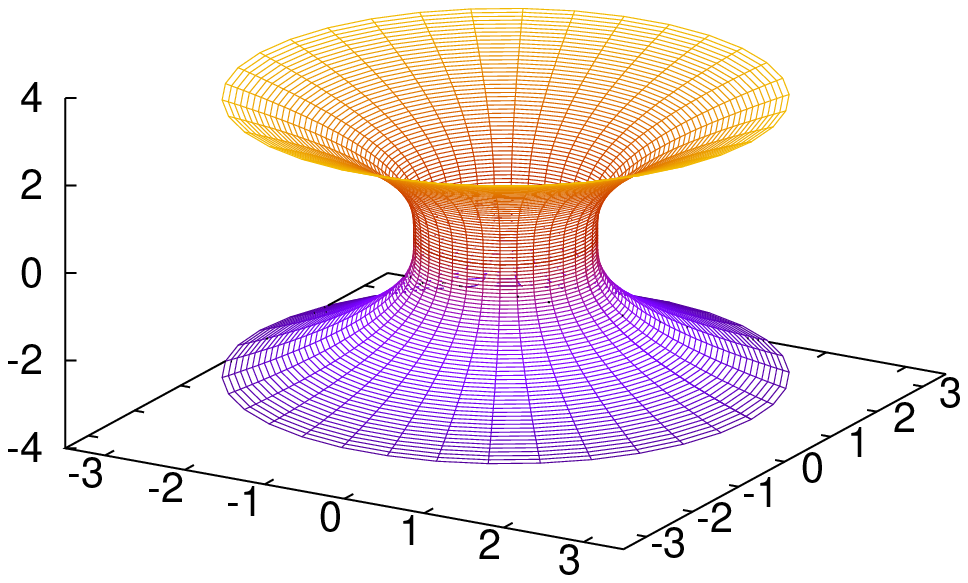}
\label{fig3a}
}
\hspace*{-2.8cm}
\subfigure[][]{
\includegraphics[height=.33\textheight, angle =0]{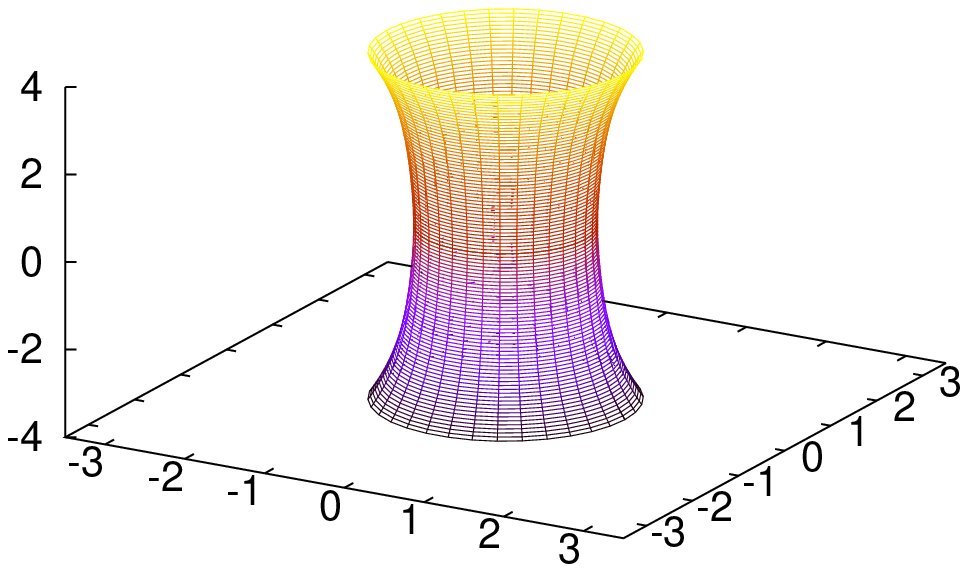}
\label{fig3b}
}
}
\end{center}
\vspace*{-0.5cm}
\caption{\small
Embedding diagrams of the wormhole throat are shown for 
solutions with $k=1$ (a)
and $k=6$ (b), and $ \alpha=1$.
}
\label{fig3}
\end{figure}

We exhibit in Fig.~\ref{fig2} the metric functions $A(x)$ and $N(x)$
for the wormhole solutions with $k$ nodes, $k=1,...,6$,
and with $ \alpha=1$.
Both metric functions are symmetric functions.
At $x=0$, 
the metric function $N(x)$ always assumes its maximal value of one.
The two symmetrically located minima of $N$ 
decrease with increasing node number $k$.
The metric function $A(x)$ has its single minimum at $x=0$.
For large node numbers, $A(x)$ is a rather flat function
in the vicinity of the throat.

To visualize the shape of the wormholes, we exhibit in Fig.~\ref{fig3}
embedding diagrams, showing the throat region
for the $k=1$ wormhole in \ref{fig3a}
and for the $k=6$ wormhole in \ref{fig3b}. 
While the areal radius $R$ and thus the area of the throat 
${\cal A}_{\rm th}$ is the same for both, 
${\cal A}_{\rm th} = 4 \pi \eta_0^2$,
the throat region becomes elongated with increasing node number $k$,
i.e., close to the throat
$z(\eta)$ increases faster with $\eta$ for the larger $k$.

\boldmath
\subsection{Wormhole properties}
\unboldmath



\begin{figure}[t!]
\begin{center}
\mbox{\hspace*{-1.0cm}
\subfigure[][]{
\includegraphics[height=.27\textheight, angle =0]{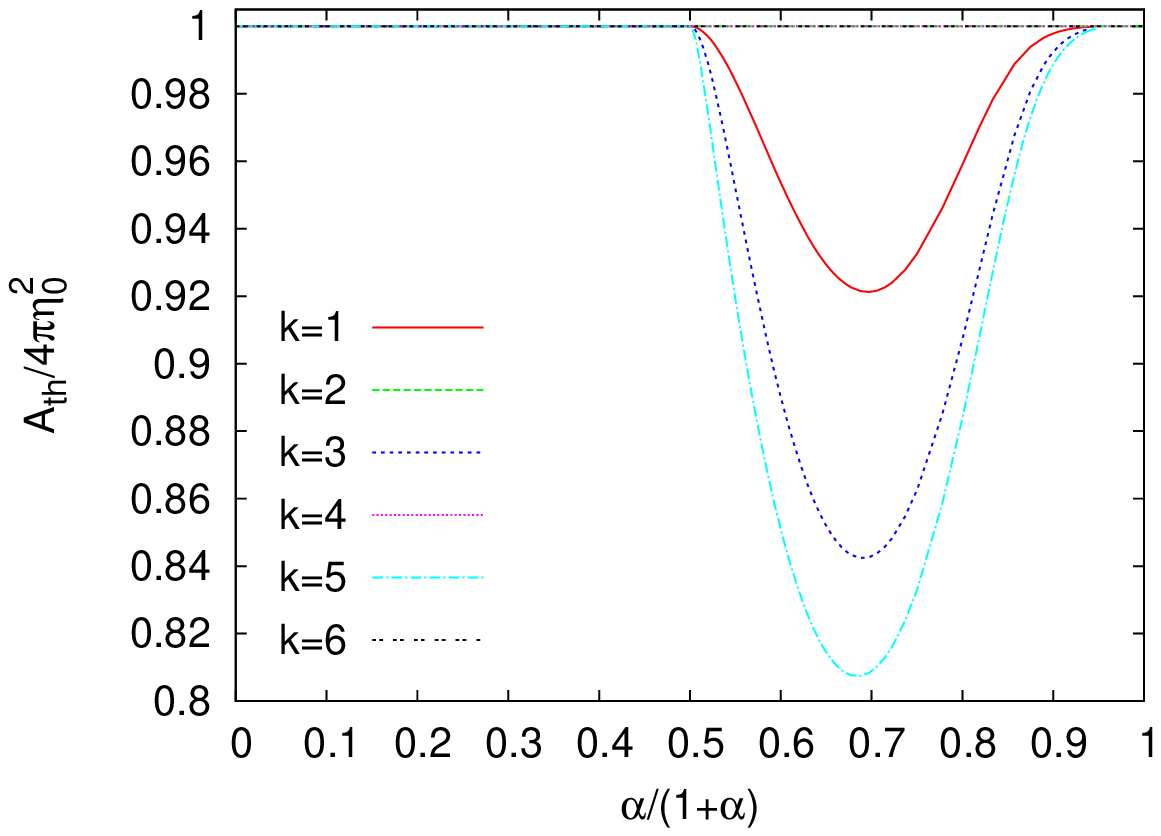}
\label{fig5a}
}
\hspace*{-1.0cm}
\subfigure[][]{
\includegraphics[height=.27\textheight, angle =0]{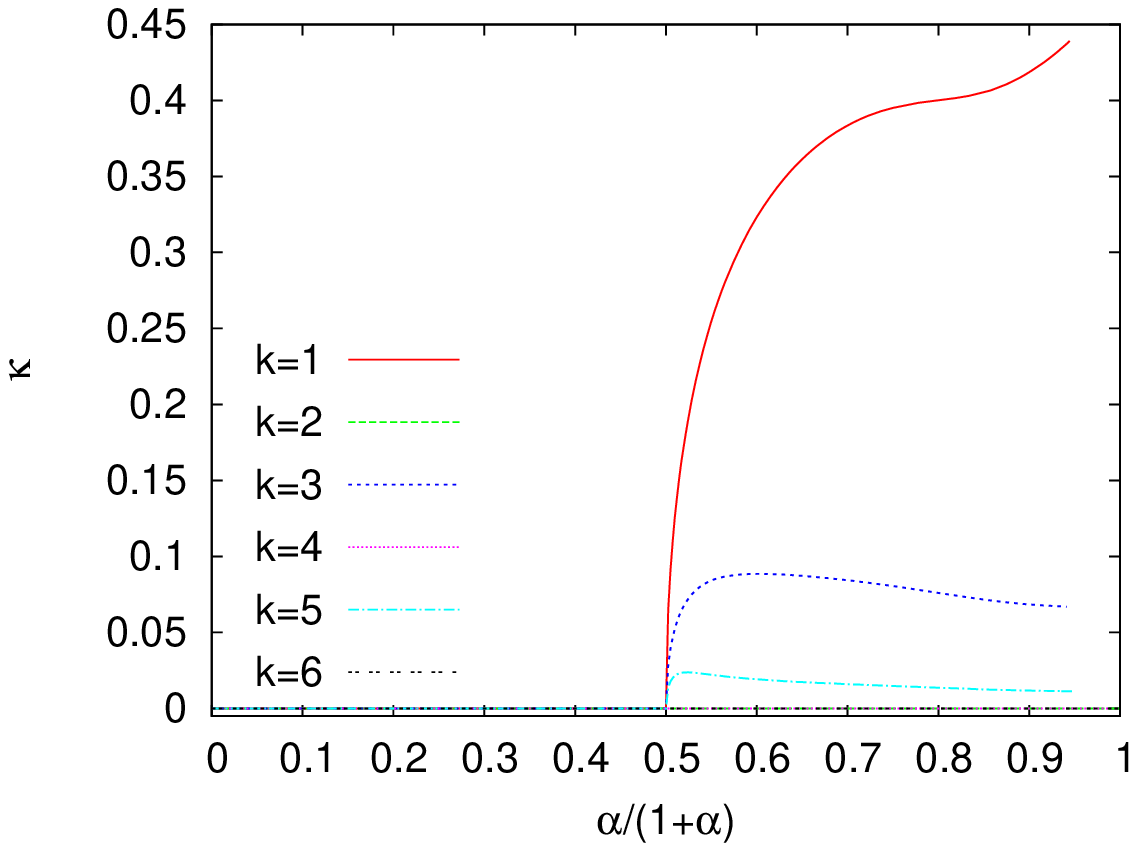}
\label{fig5b}
}
}
\end{center}
\vspace*{-0.5cm}
\caption{\small
Wormhole properties are shown versus $ \alpha$ for
wormhole solutions with $k=1,...,6$:
the scaled area of the throat ${\cal A}_{\rm th}/4 \pi \eta_0^2$ (a),
and the surface gravity of the throat $\kappa$ (b).
}
\label{fig5}
\end{figure}

Let us now consider the physical
properties of the wormhole solutions.
Physical quantities of interest at the throat are the 
area of the throat ${\cal A}_{\rm th}$ and 
the surface gravity at the throat $\kappa$.
To be able to exhibit these quantites over the full range of
the coupling constant $ \alpha$,
we here consider the dependence on the compactified coupling constant
$ \alpha/(1+ \alpha)$.

We exhibit in Fig.~\ref{fig5a} the 
scaled throat area ${\cal A}_{\rm th}/4 \pi \eta_0^2$.
Clearly, the scaled throat area remains constant
as long as the throat is located at $\eta=0$.
In this respect we need to consider even-$k$ and odd-$k$ solutions separately.
For even-$k$ wormholes 
$ \alpha_{\rm cr}$ increases faster than $ \alpha$.
Thus even-$k$ wormholes always possess only a single throat,
located at $\eta=0$.
Consequently, their scaled throat area is constant throughout, 
${\cal A}_{\rm th}/4 \pi \eta_0^2 = 1$.

In contrast, for odd-$k$ wormholes the throat is only located at $\eta=0$
as long as $ \alpha$ does not exceed their common critical value
$ \alpha_{\rm cr}=1$. Beyond $ \alpha_{\rm cr}$
an equator is located at $\eta=0$ 
and the scaled area of each of the two throats -
being minimal area surfaces - decreases with increasing $ \alpha$.
This decrease is the stronger the larger the node number $k$.
At a value of $ \alpha \approx 3.5$, 
which is rather independent of the odd node number,
the scaled throat area reaches a minimum
as a function of $ \alpha$. 
Subsequently it increases again towards 
${\cal A}_{\rm th}/4 \pi \eta_0^2 =1$, as $ \alpha \to \infty$.

\begin{figure}[t!]
\begin{center}
\mbox{\hspace*{-1.0cm}
\subfigure[][]{
\includegraphics[height=.30\textheight, angle =0]{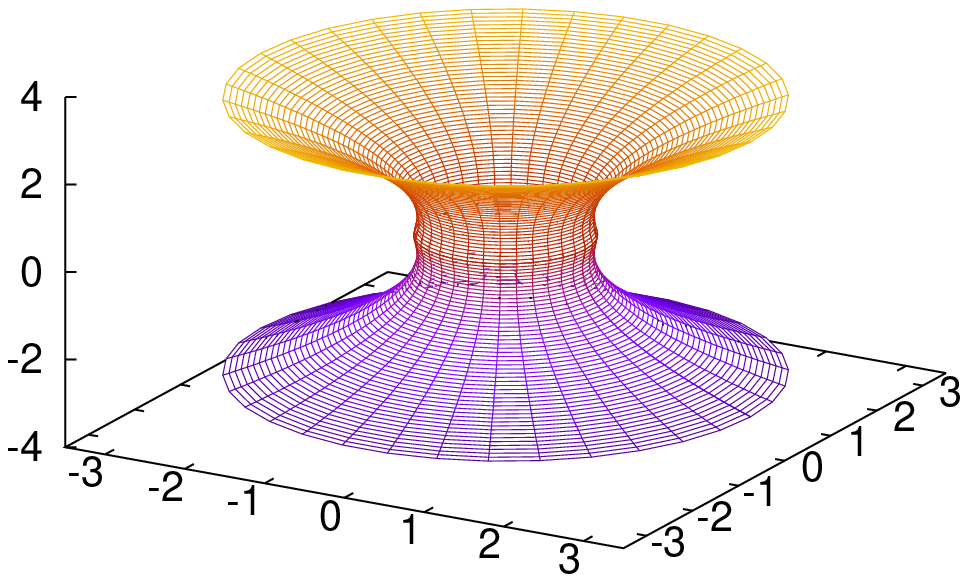}
\label{fig6a}
}
\hspace*{-1.0cm}
\subfigure[][]{
\includegraphics[height=.30\textheight, angle =0]{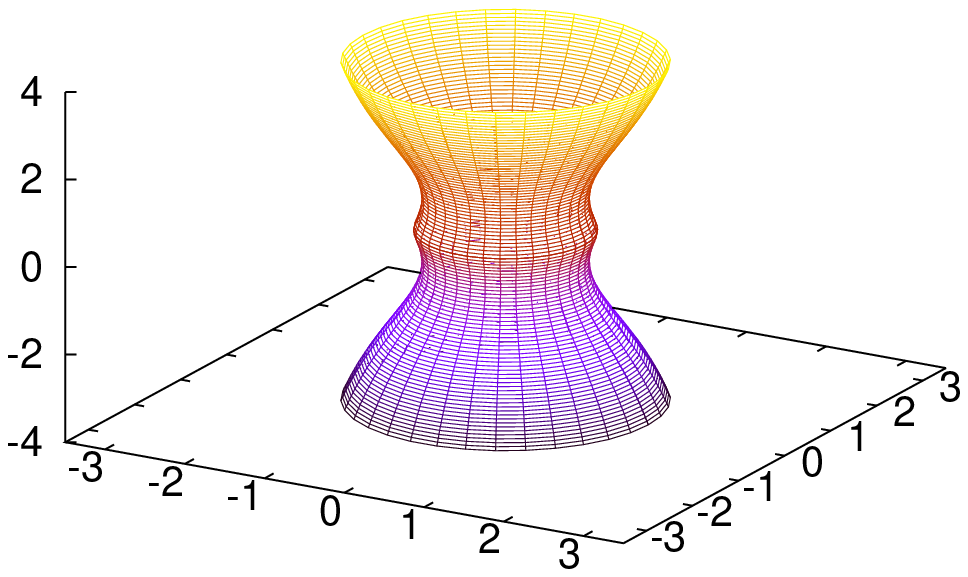}
\label{fig6b}
}
}
\end{center}
\vspace*{-0.5cm}
\caption{\small
Embedding diagrams of double-throat wormholes are shown for $k=1$ (a)
and $k=5$ (b), and $ \alpha=3$.
}
\label{fig6}
\end{figure}

The surface gravity $\kappa$ at the throat is shown in Fig.~\ref{fig5b}.
It vanishes for all even-$k$ solutions,
since even-$k$ wormholes possess only a single throat.
However, for odd-$k$ wormholes $\kappa$ increases from zero
as soon as the double-throat appears.
Thus $\kappa$ has finite values beyond $ \alpha_{\rm cr}=1$.
Here the surface gravity is the smaller the higher the odd node number.
We exhibit two examples of embedding
diagrams of double-throat wormholes in Fig.~\ref{fig6}.

\begin{figure}[t!]
\begin{center}
\mbox{\hspace*{-1.0cm}
\subfigure[][]{
\includegraphics[height=.27\textheight, angle =0]{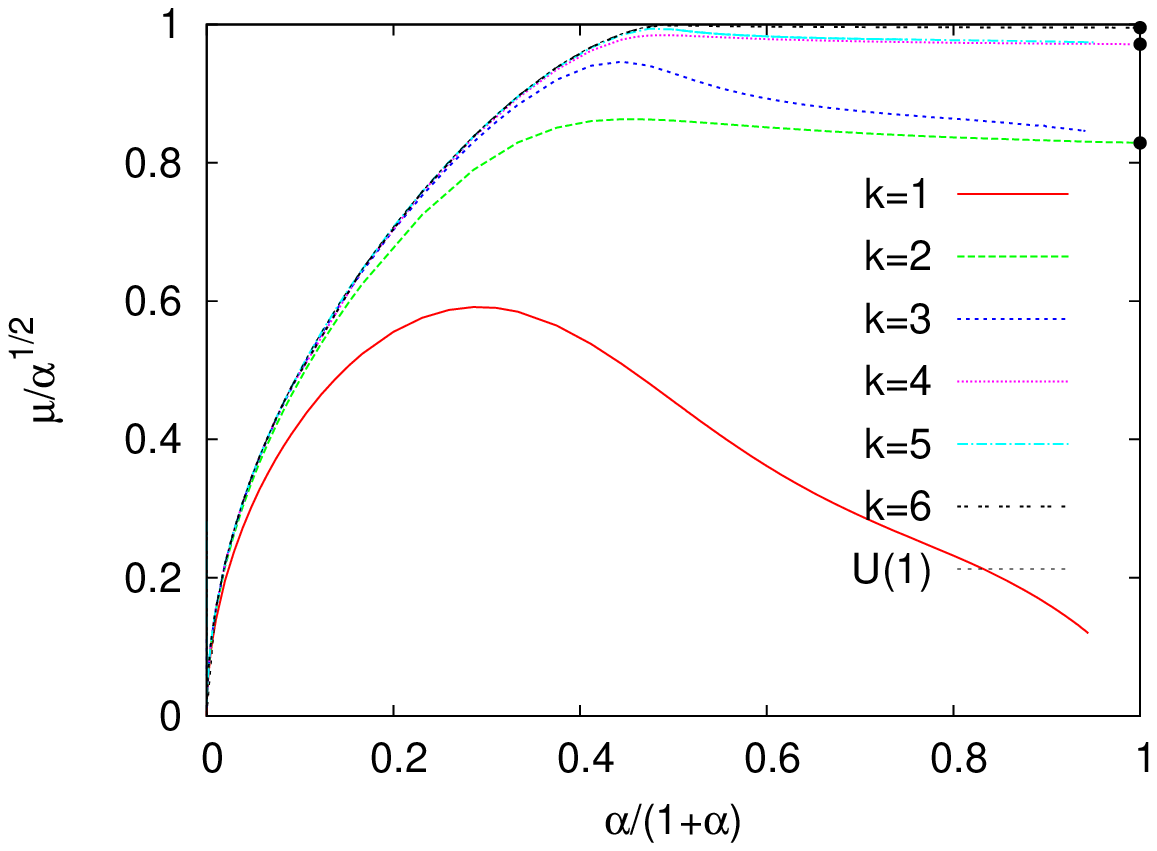}
\label{fig7a}
}
\hspace*{-1.0cm}
\subfigure[][]{
\includegraphics[height=.27\textheight, angle =0]{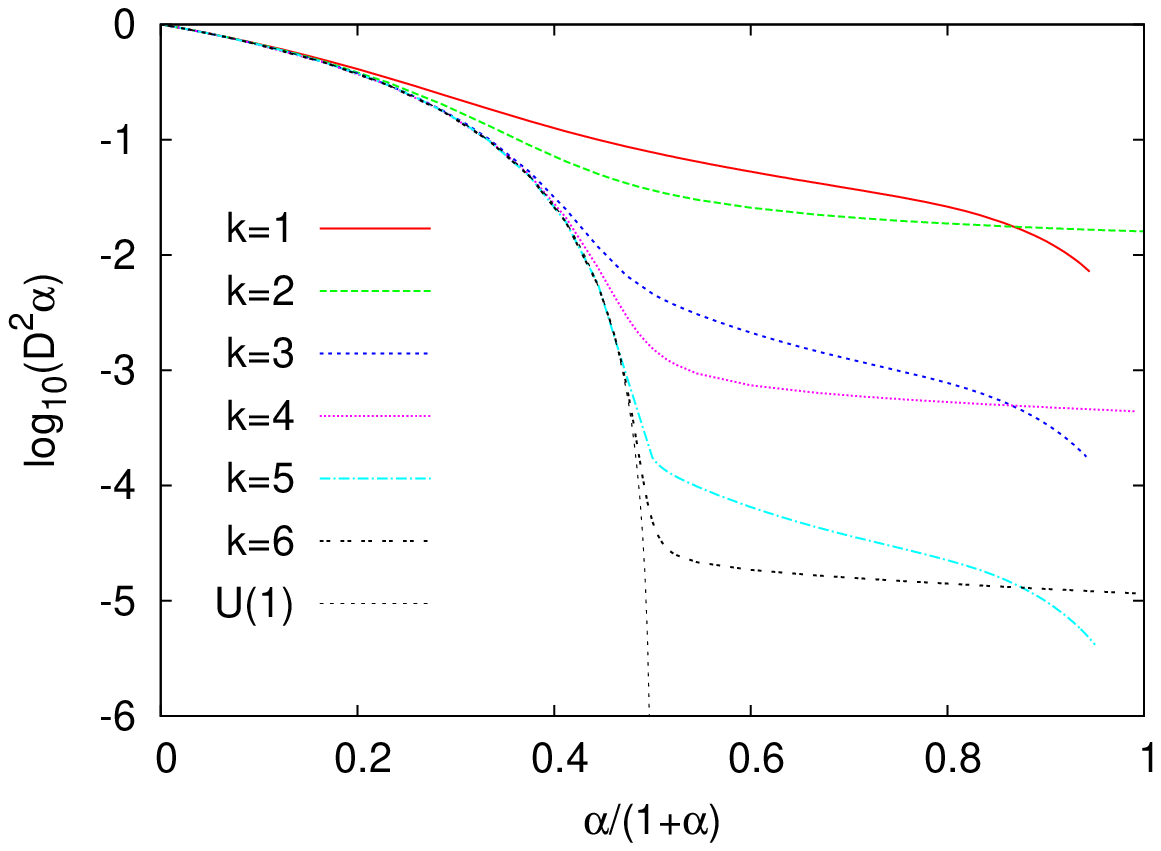}
\label{fig7b}
}
}
\end{center}
\vspace*{-0.5cm}
\caption{\small
Wormhole properties are shown versus $ \alpha$ for
wormhole solutions with $k=1,...,6$:
the scaled mass parameter $\mu/\sqrt{ \alpha}$ (a) 
and the scaled scalar charge $D \sqrt{ \alpha}$ (b).
The dots in (a) correspond to the masses of
the Bartnik-McKinnon solutions with $n=k/2$ nodes (for even $k$).
Also shown are the scaled mass and the scaled scalar charge
for the embedded U(1) wormholes.
}
\label{fig7}
\end{figure}

Let us now turn to the global charges of the wormholes.
Starting from the zero mass limit of
the Ellis wormholes, the mass parameter $\mu$ of the 
wormhole solutions increases with increasing $ \alpha$.
For large $ \alpha$ this increase is roughly proportional to
$\sqrt{ \alpha}$, except for the lowest mass solution with $k=1$.
We therefore exhibit the scaled mass $\mu/\sqrt{ \alpha}$
versus the compactified coupling constant
$ \alpha/(1+ \alpha)$ in Fig.~\ref{fig7a},
presenting solutions with node numbers $k=1,...,6$.

The figure shows, that for a fixed node number $k>1$
the scaled mass $\mu/\sqrt{ \alpha}$
reaches a finite limiting value as $ \alpha \to \infty$.
Only for $k=1$ the scaled mass appears to tend to
zero in this limit.
Moreover, the limiting value of an odd-$k$ solution with $k=2l+1$ nodes
seems to agree with the limiting value of an even-$k$ solution with $k=2l$ nodes.
We note, that we have to extrapolate this mass for the odd-$k$ solutions 
in the limit $ \alpha \to \infty$,
since the numerical calculations encounter difficulties in this case. 

To address this problem, let us consider the behavior of the even-$k$ 
and the odd-$k$ solutions at the center, $\eta=0$.
For odd-$k$ wormholes,
the derivative of the gauge field function $K$ increases
strongly with increasing $ \alpha$ at $\eta=0$.
At the same time, the minimum of
the metric function $A$ decreases sharply towards zero at $\eta=0$.
This indicates, that a singular behavior should be encountered at $\eta=0$
in the limit $ \alpha \to \infty$ (see the discussion below in 
section \ref{limit1}).

For even-$k$ wormholes, on the other hand,
the gauge field function $K$ increases slowly with increasing $ \alpha$
at $\eta=0$, while the metric function $A$ decreases slowly with
increasing $ \alpha$ at $\eta=0$, and broadens at the same time.
Thus, in contrast to the odd-$k$ case, 
here a smooth limit $ \alpha \to \infty$ is observed
(see the discussion below in section \ref{limit1}).

The scalar charge $D$ is considered for the same set of solutions in Fig.~\ref{fig7b}.
For large $ \alpha$, the scalar charge tends to zero.
In order to identify the limiting behavior, we exhibit
the scaled scalar charge $D \sqrt{ \alpha}$ in the figure.
The scaled scalar charge of an even-$k$ solution approaches a finite limiting value,
when $ \alpha \to \infty$.
The limiting value of the scaled scalar charge of an odd-$k$ solution 
is hard to estimate, however.

\boldmath
\subsection{Limit $ \alpha \to \infty$}\label{limit1}
\unboldmath

\begin{figure}[p!]
\begin{center}
\mbox{\hspace*{-1.0cm}
\subfigure[][]{
\includegraphics[height=.27\textheight, angle =0]{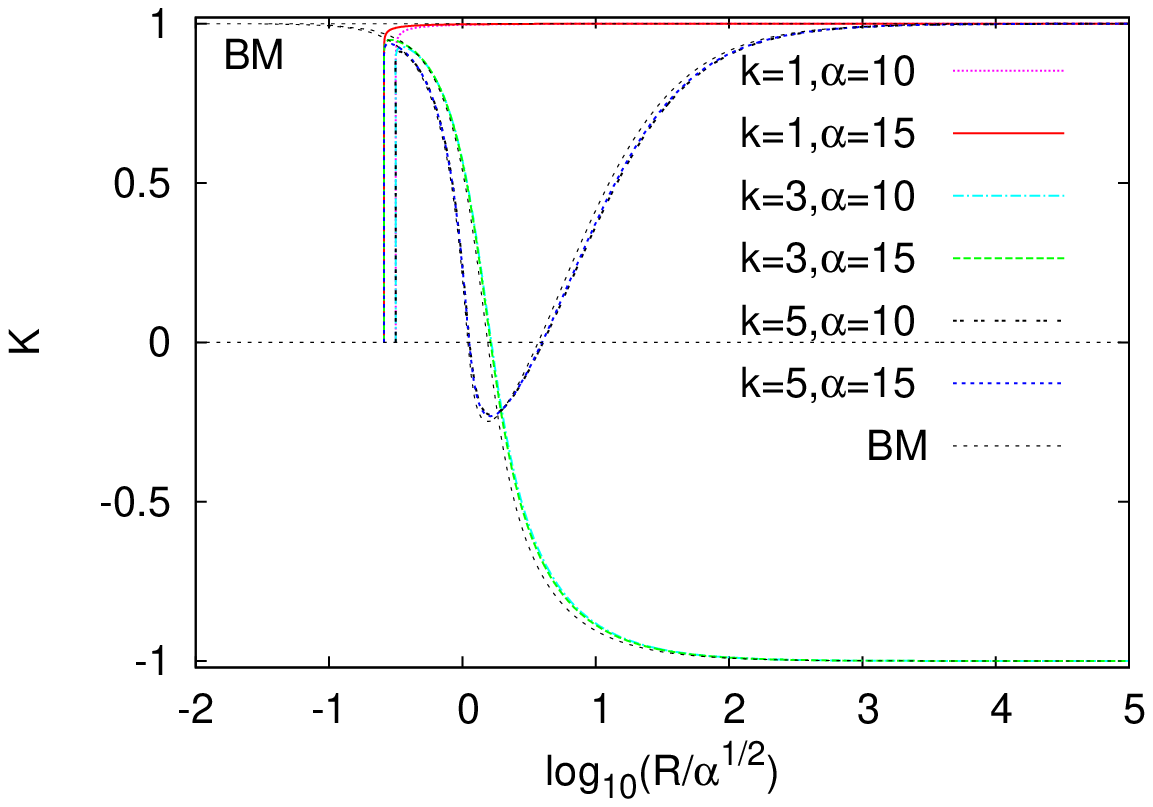}
\label{fig8a}
}
\hspace*{-1.0cm}
\subfigure[][]{
\includegraphics[height=.27\textheight, angle =0]{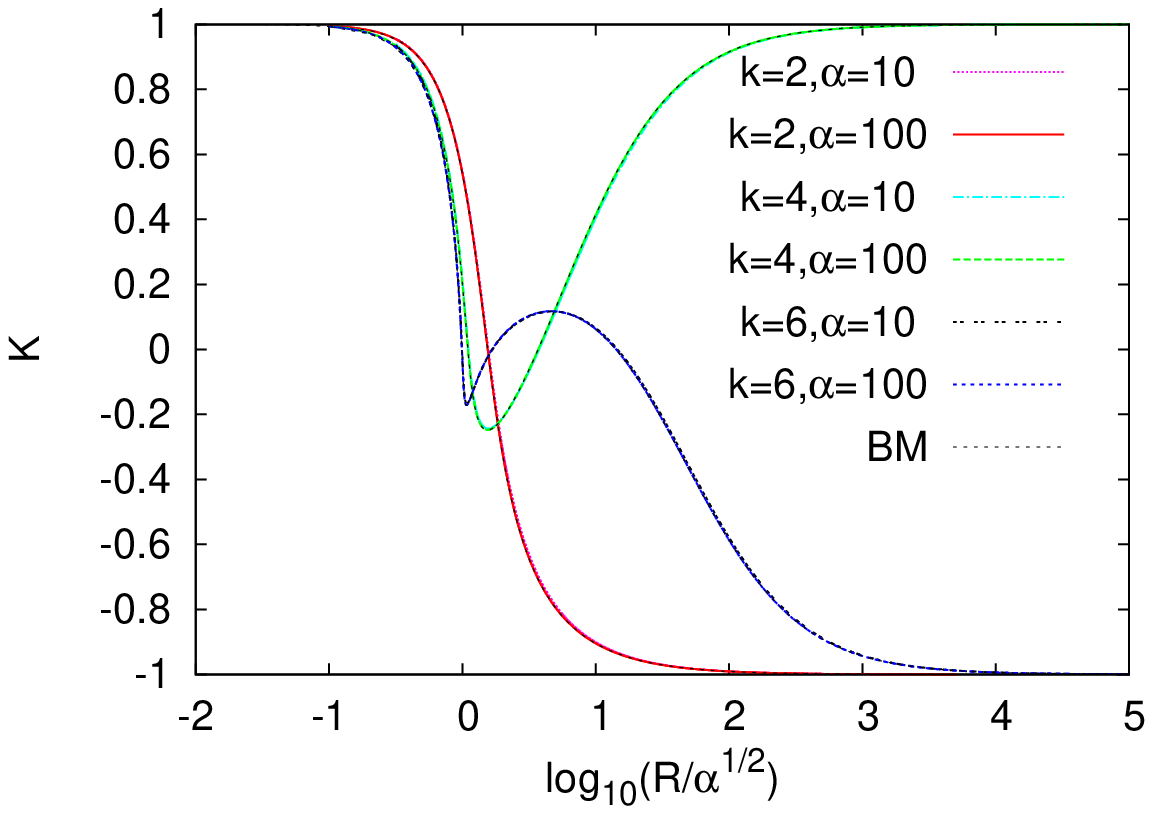}
\label{fig8b}
}
}
\mbox{\hspace*{-1.0cm}
\subfigure[][]{
\includegraphics[height=.27\textheight, angle =0]{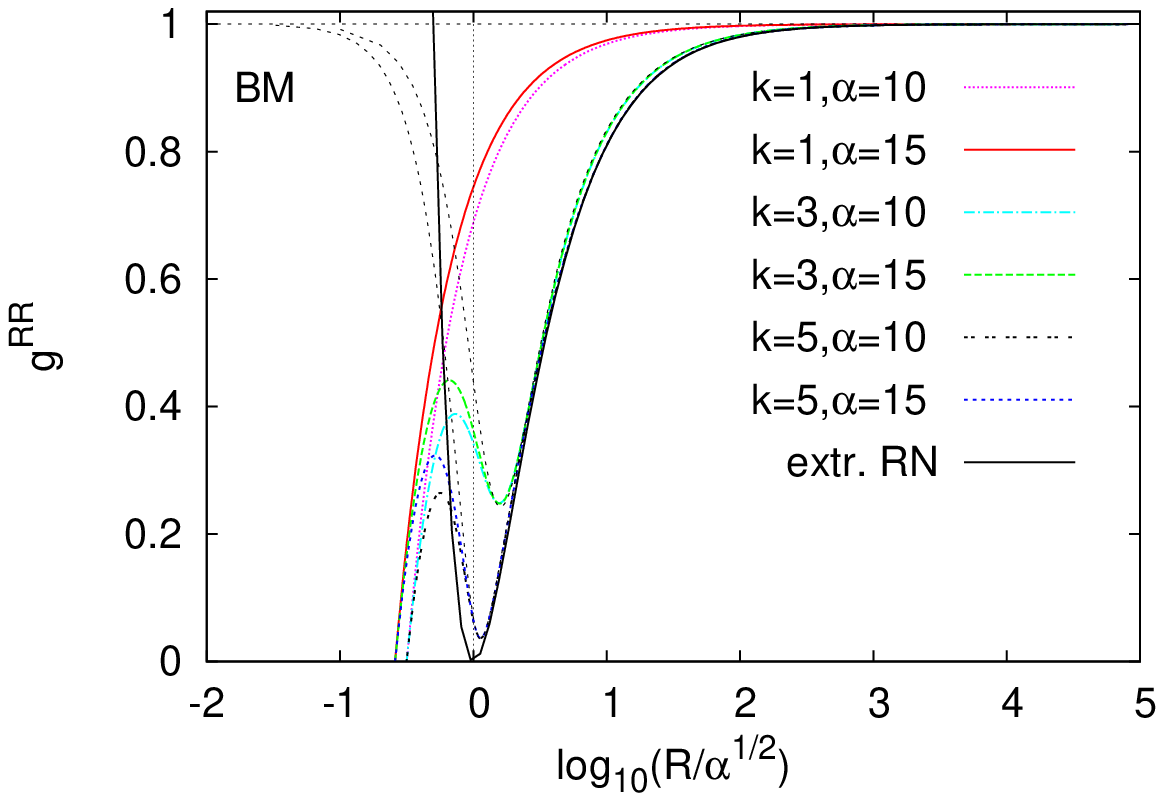}
\label{fig8c}
}
\hspace*{-1.0cm}
\subfigure[][]{
\includegraphics[height=.27\textheight, angle =0]{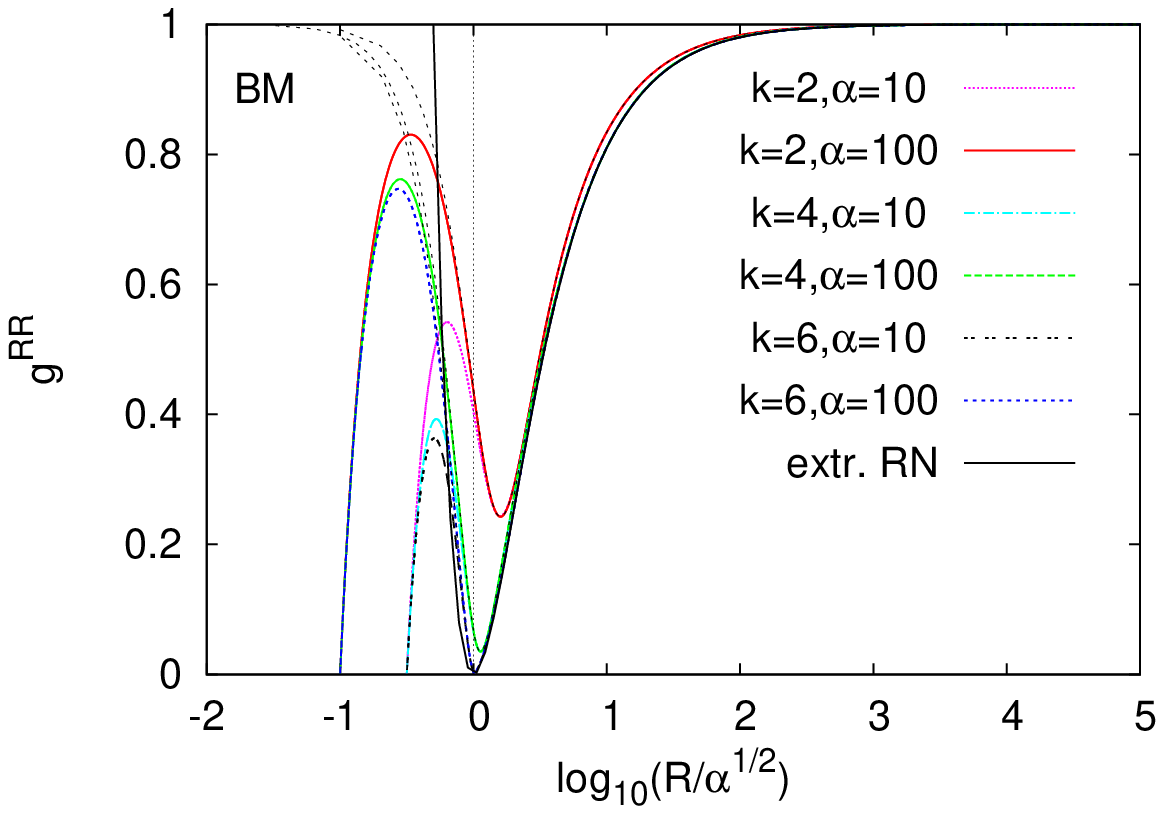}
\label{fig8d}
}
}
\mbox{\hspace*{-1.0cm}
\subfigure[][]{
\includegraphics[height=.27\textheight, angle =0]{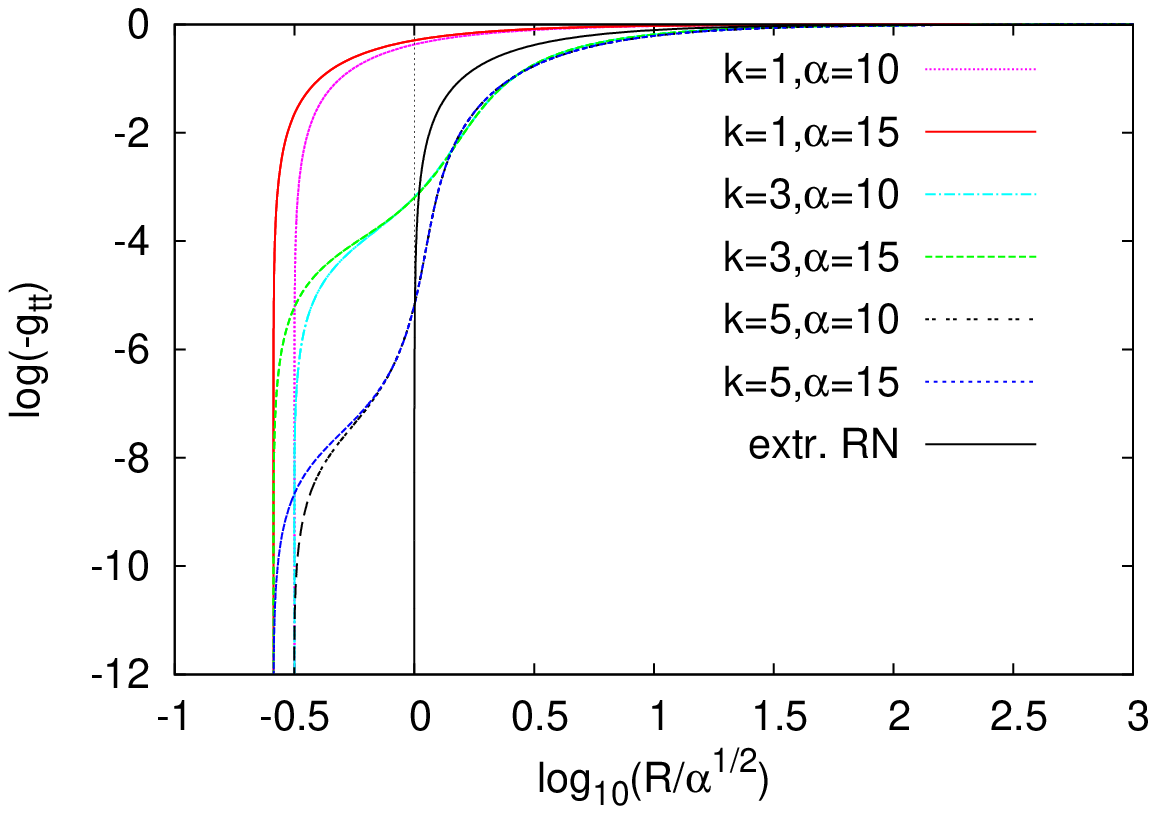}
\label{fig8e}
}
\hspace*{-1.0cm}
\subfigure[][]{
\includegraphics[height=.27\textheight, angle =0]{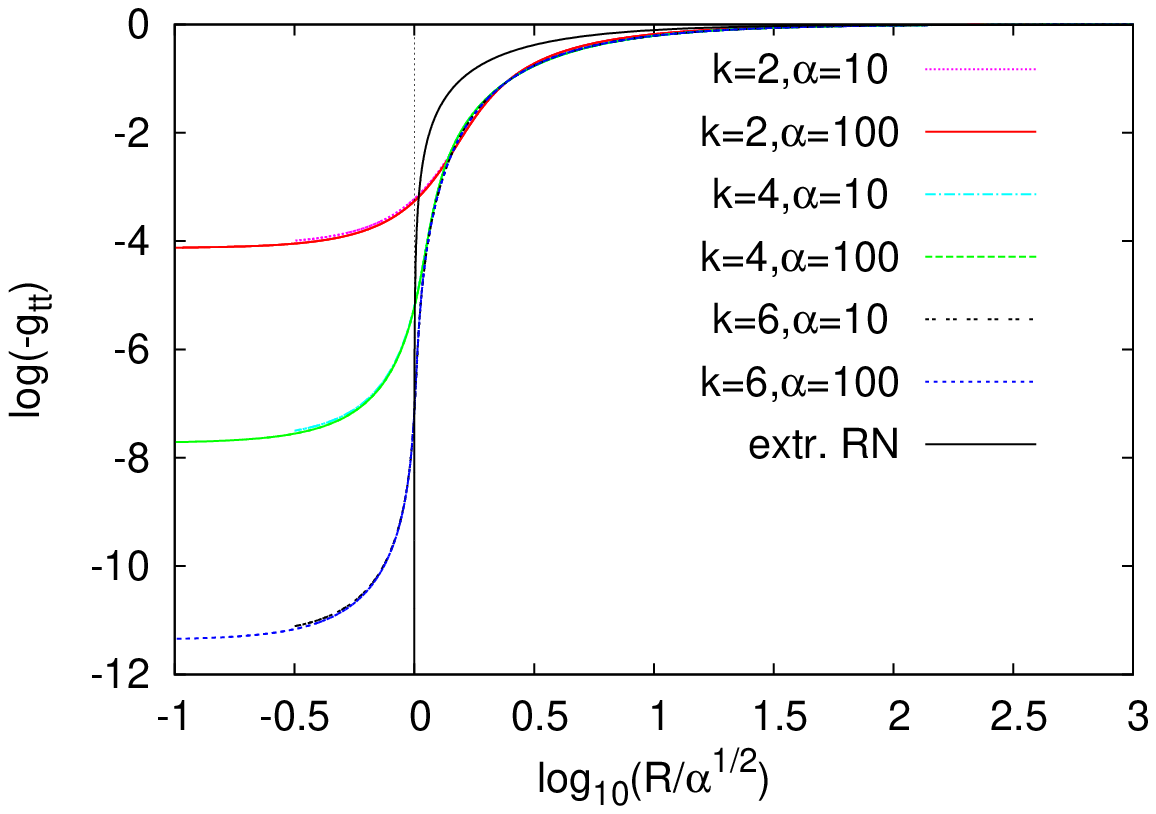}
\label{fig8f}
}
}

\end{center}
\vspace*{-0.5cm}
\caption{\small
Wormhole functions $K$, $g^{RR}$ and $g_{tt}$ are shown versus the scaled 
Schwarzschild-like coordinate $R/ \sqrt{ \alpha}$.
The left column corresponds to odd-$k$ solutions with $k=1$, 3 and 5
and $\sqrt{ \alpha}=10$ and 15;
the right column to even-$k$ solutions with $k=2$, 4, 6 and $ \alpha = 10$ and 100.
Also shown are the respective functions
of the Bartnik-McKinnon solutions with $n=1$, 2 and 3 nodes
and of the extremal Reissner-Nordstr\"om solution with unit charge.
}
\label{fig8}
\end{figure}

Let us now discuss the wormhole solutions for large 
values of $ \alpha$, considering in particular the limit $ \alpha \to \infty$.
The limit ${\alpha}=\bar \alpha/\eta_0^2 \to \infty$ can be obtained either
by keeping $\eta_0$ fixed and taking $\bar \alpha \to \infty$ or by keeping
$\bar \alpha$ fixed and taking $\eta_0 \to 0$. 
In the latter limit the wormhole throat will
shrink to zero size.

We have seen, that when scaling the mass 
of the wormhole solutions with $\sqrt{{\alpha}}$ we obtain
finite values for the scaled mass in the limit $ \alpha \to \infty$.
It therefore suggests itself to introduce the scaled coordinate
\begin{equation}
\bar{x} = x/\sqrt{{\alpha}} \ 
\end{equation}
in order to study the limit $ \alpha \to \infty$.
The ODEs Eqs.~(\ref{eqN})-(\ref{eqK})
then read
\begin{eqnarray}
N'' & = & 
-2\frac{(K^2-1)^2}{\bar{f}^2 N} 
+  \frac{2 A(1- N -2 \bar{x} N') 
- A'( \bar{f}N' - 2 \bar{x} N)}{A \bar{f}} \ ,
\label{eqNb}\\
A''& = & 
A \frac{2N\bar{f}K'^2 + (K^2-1)^2}{\bar{f}^2 N^2} 
-A'\frac{\bar{f} N' + 2\bar{x} N}{\bar{f}N} \ ,
\label{eqAb}\\
K'' & = &  \frac{A K(K^2-1) - N A'\bar{f} K'}{AN\bar{f}} \ 
\label{eqKb}
\end{eqnarray}
with $\bar{f} = \bar{x}^2 +1/{\alpha}$.

In the limit ${\alpha}\to \infty$ we observe that $\bar{f} = \bar{x}^2$
and the ODEs Eqs.~(\ref{eqNb})-(\ref{eqKb}) reduce to the
Einstein-Yang-Mills equations. However, only for 
even-$k$ wormhole solutions
the boundary condition $K'(0)=0$ is consistent with the
regularity condition $K(0)=\pm 1$ of the Einstein-Yang-Mills solutions.
For odd-$k$ wormholes the boundary condition $K(0)=0$ must be met,
which is in conflict with the regularity condition.

The gauge field functions and the metric functions 
of wormhole solutions with $k$ nodes, $k=1,...,6$, 
are exhibited in Fig.~\ref{fig8} in the limit of large $ \alpha$.
Here the left set of figures corresponds to the odd-$k$  solutions,
and the right set of figures to the even-$k$ ones.
To better demonstrate the limit, we here consider the 
Schwarzschild-like areal coordinate $R$, Eq.~(\ref{R}),
and employ the scaled areal coordinate $R/\sqrt{ \alpha}$
in the figures.

The even-$k$ wormhole solutions 
converge smoothly to the regular Bartnik-McKinnon solutions
in the limit  $ \alpha \to \infty$.
This is demonstrated in Fig.~\ref{fig8b} for the gauge field function $K$
of the solutions with $2$, $4$ and $6$ nodes
and the values of $ \alpha=10$ and 100, 
where the corresponding limiting Bartnik-McKinnon solutions
are also shown for comparison.

Since the wormhole throat decreases to zero size in the limit $ \alpha \to \infty$,
the spacetime splits into two disconnected 
asymptotically flat spacetimes in that limit.
In each one of them a Bartnik-McKinnon solution is found,
which has precisely $n=k/2$ nodes.
Consequently, the scaled mass
$\mu/\sqrt{ \alpha}$ of the limiting solution 
obtained for even node number $k$
agrees with the mass of the
respective Bartnik-McKinnon solution with node number $n=k/2$,
as seen in Fig.~\ref{fig7a}.

In Fig.~\ref{fig8d} the metric function $g^{RR}$ is depicted
for this set of solutions. 
$g^{RR}$ is obtained by transforming the radial coordinate
to the Schwarzschild-like radial coordinate $R$.
The convergence towards the respective Bartnik-McKinnon solutions is clearly seen
for these large values of $ \alpha$.
Deviations from the limiting solutions occur in the small region
close to the throat.
Fig.~\ref{fig8f} exhibits the convergence analogously
for the metric function $g_{tt}$ in Schwarzschild-like coordinates.

For the odd-$k$ solutions the situation is different, because
of the incompatibility of the boundary conditions at $\eta=0$.
This is demonstrated in Fig.~\ref{fig8a} for the gauge field function $K$
of the solutions with $1$, $3$ and $5$ nodes
and values of $ \alpha=10$ and 15.
Here the gauge field function must assume the value zero at the throat. 
This leads to a sudden sharp decrease of the functions near the throat.
It is clear, that the huge derivatives associated with such a sharp decrease
imply a severe numerical challenge.

However, away from the vicinity of the throat the higher odd-$k$ solutions also 
approach the Bartnik-McKinnon solutions, as seen in the figure.
In particular, the $k=3$ solution appoaches the $n=1$ 
Bartnik-McKinnon solution, and the $k=5$ solution the $n=2$
Bartnik-McKinnon solution.
This is in accord with the limiting behavior of the masses,
as displayed in Fig.~\ref{fig7a}.

The convergence of the odd-$k$ solutions outside the throat region towards
the Bartnik-McKinnon solutions is also seen 
in Figs.~\ref{fig8c} and \ref{fig8e},
where the metric functions $g^{RR}$ and $g_{tt}$ are depicted
in Schwarzschild-like coordinates.
Inspecting the behavior of the wormhole solutions
for odd-$k$ further, we conclude, that singularities indeed appear in the limit.
With increasing $ \alpha$ the Kretschmann scalar increases
dramatically at $\eta=0$, indicating that a curvature singularity
is approached.

\boldmath
\subsection{Limit $k \to \infty$}
\unboldmath

\begin{figure}[t!]
\begin{center}
\mbox{\hspace*{-1.0cm}
\subfigure[][]{
\includegraphics[height=.27\textheight, angle =0]{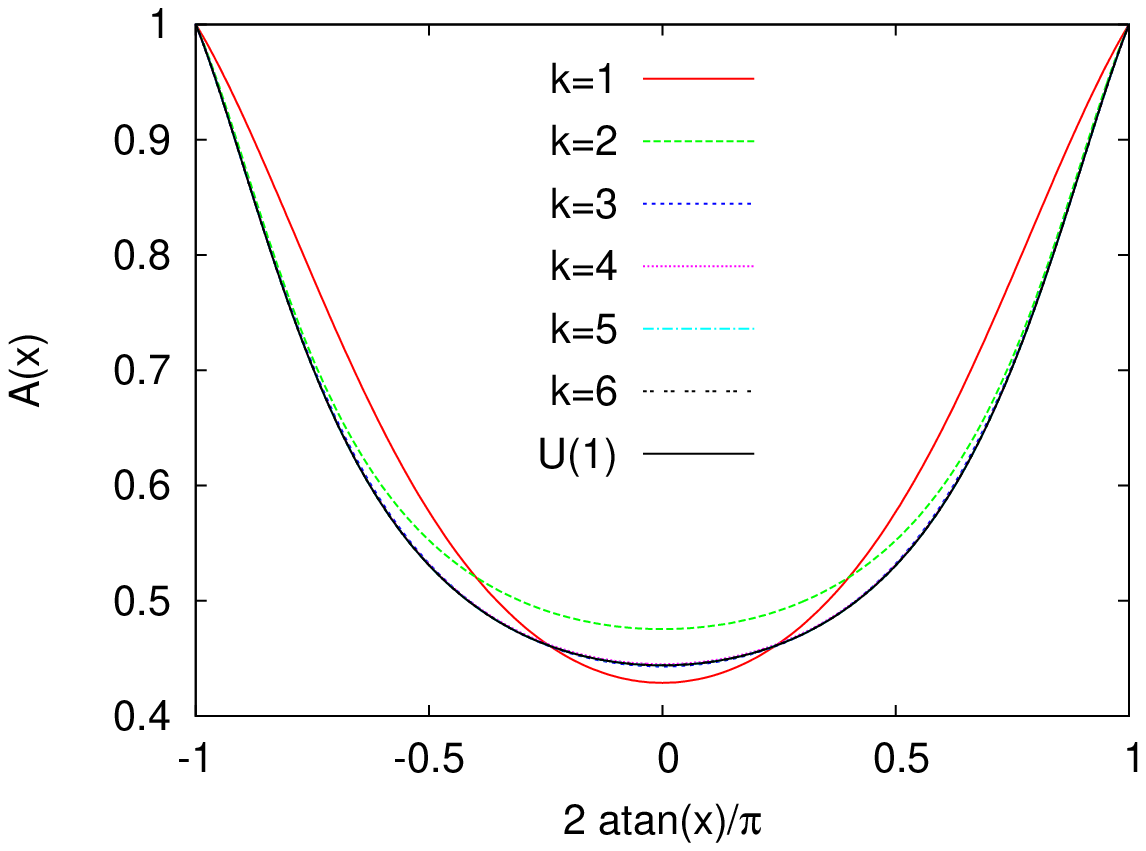}
\label{fig9a}
}
\hspace*{-1.0cm}
\subfigure[][]{
\includegraphics[height=.27\textheight, angle =0]{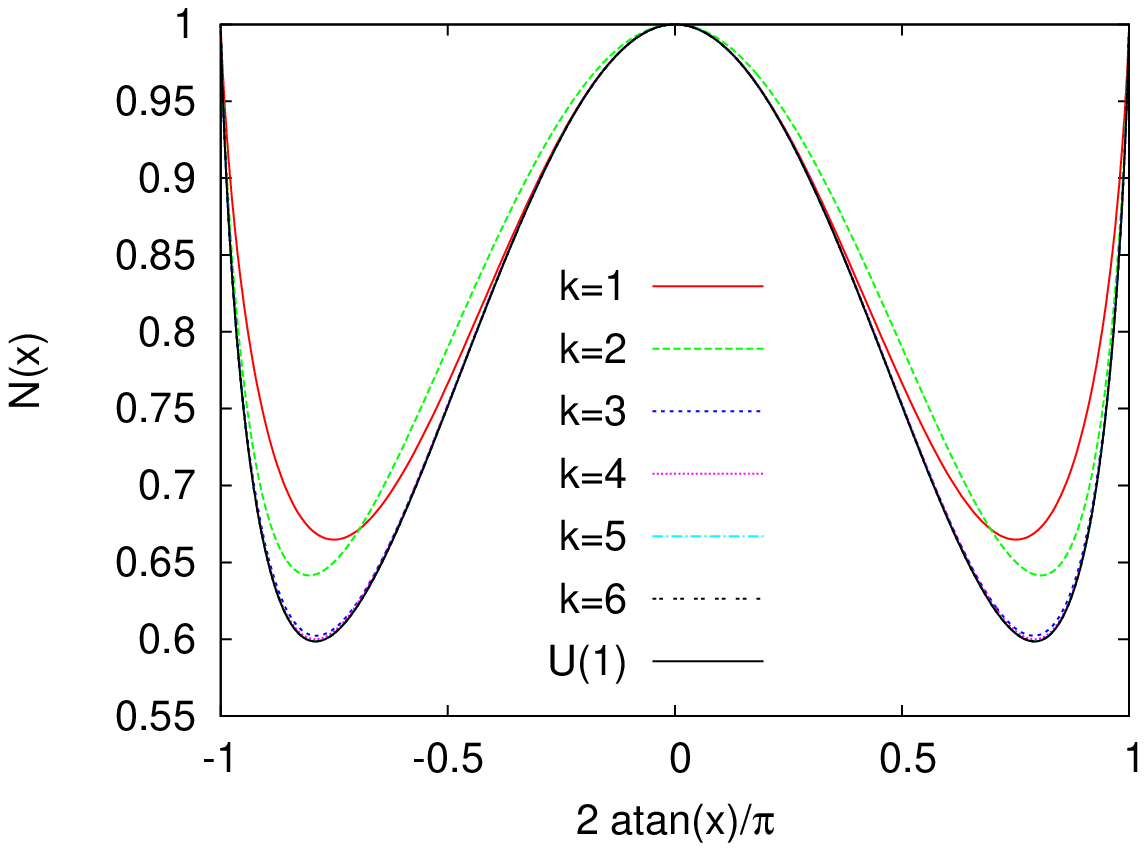}
\label{fig9b}
}
}
\end{center}
\vspace*{-0.5cm}
\caption{\small
Wormhole functions $A$ (a) and $N$ (b) are shown for solutions with $k$ nodes, $k=1,...,6$,
and $ \alpha = 0.5$.
The corresponding functions of the limiting Abelian solution are also shown.
}
\label{fig9}
\end{figure}

Let us finally address the limiting behavior of the wormhole solutions
for fixed values of $ \alpha$, when the node number $k$ increases,
$k \to \infty$.
Here we must distinguish two cases.
When $ \alpha < 1$, a U(1) wormhole solution with unit charge exists,
that can be embedded into SU(2),
as discussed earlier in section \ref{exact}.
In contrast, for $ \alpha > 1$ there is no such embedded wormhole solution.

For fixed $ \alpha$ and  increasing node number $k$,
the non-Abelian wormhole solutions
converge towards the corresponding Abelian
solution with the same $ \alpha$,
as long as $ \alpha < 1$.
This convergence is demonstrated in Fig.~\ref{fig9} for
the metric functions $A$ and $N$ of the wormhole solutions
with $k$ nodes, $k=1,...,6$
and $ \alpha=0.5$.
For small values of $ \alpha$ this convergence is very fast,
as seen in the figure.
At $ \alpha =1$ the Abelian solution becomes singular in these coordinates.

To understand the limiting behavior of the wormhole solutions 
for $ \alpha > 1$,
let us recall the limiting behavior of the Bartnik-McKinnon solutions.
With increasing node number $n$, $n\to \infty$,
these approach the embedded extremal 
Abelian Reissner-Nordstr\"om black hole solution
with unit charge in an outer region $R>1$, where $R$ is the Schwarzschild-like
radial coordinate.
However, in the inner region $R<1$ the Bartnik-McKinnon solutions approach
a non-Abelian limiting solution.

Let us therefore look at the behavior of the 
non-Abelian wormhole solutions for $ \alpha > 1$
in Schwarzschild-like coordinates, 
and compare their limiting behavior
for $k \to \infty$ with the one of the Bartnik-McKinnon solutions.
The Schwarzschild-like metric functions $g_{RR}$ and $g_{tt}$
of the wormhole solutions with node numbers $k=1,...,6$
and several values of $ \alpha >1$
are exhibited in Fig.~\ref{fig8} in these coordinates.

These figures also contain the respective metric functions of the
extremal Reissner-Nordstr\"om solution with unit charge.
We observe, that  the metric functions
of the wormholes converge towards the metric functions of the
extremal Reissner-Nordstr\"om solution 
in the outer region $R/\sqrt{ \alpha}>1$.
This is not surprising, since in this region
the wormhole solutions converge towards the
Bartnik-McKinnon solutions, which themselves
converge to the extremal Abelian Reissner-Nordstr\"om solution
with unit charge.

In the inner region $R/\sqrt{ \alpha}<1$
two different limiting solutions are approached 
for a fixed value of $ \alpha$.
One limiting solution is associated with the even-$k$ wormholes 
and another one with the odd-$k$ wormholes.
These limiting solutions are genuinely non-Abelian.
They differ from each other and from the limiting non-Abelian
solution of the Bartnik-McKinnon solutions
close to the throat.

For the mass of the limiting solutions such a distinction is not necessary, however,
since only the asymptotic behavior is relevant.
As seen in Fig.~\ref{fig7a},
the limiting scaled mass of the wormhole solutions obtained for $k \to \infty$
corresponds to the scaled mass of the respective Abelian solution when $ \alpha < 1$,
and to unity when $ \alpha > 1$. At $ \alpha = 1$ the mass is continuous.
Thus the extremal Reissner-Nordstr\"om solution plays an as  prominent role
for the non-Abelian wormholes as for the globally regular non-Abelian solutions
and their associated hairy black holes.
Similarly, Fig.~\ref{fig7b} shows that
the limiting value of the scaled scalar charge
corresponds to the scaled scalar charge of the respective Abelian 
solution when $ \alpha < 1$. The scaled scalar charge
vanishes when $ \alpha > 1$, in accordance with the fact
that the extremal Reissner-Nordstr\"om solution has no scalar charge.

\section{Conclusions}

We have investigated 
Einstein-Yang-Mills wormholes supported by a phantom field.
The solutions are spherically symmetric and possess two asymptotically
flat regions. The non-Abelian gauge field is described in terms
of the Wu-Yang Ansatz which contains a single function $K$.
For $K=1$ the gauge field is trivial and the Ellis wormhole
is recovered, whereas for $K=0$ a set of embedded Abelian
wormhole solutions is obtained.

Our main interest here has focussed on genuine
non-Abelian wormhole solutions, where the gauge field function $K$
possesses a number of nodes, which we have labelled by the integer $k$.
Already in the probe limit, where the gauge field decouples from
the metric, a sequence of non-trivial solutions with $k$ nodes is found.
The background metric of these solutions is the metric of the Ellis wormhole.
We recall, that in the background of a flat spacetime no 
such non-trivial non-Abelian solutions are possible.

In order to take the backreaction into account, the coupled set of Einstein-Yang-Mills-scalar
equations has been solved. 
For a given value of the throat size and a given node number $k$,
a family of wormhole solutions emerges from the solution
obtained in the probe limit,
as the coupling to gravity is increased from zero.
This family of solutions changes smoothly with the coupling constant.
We consider these solutions as wormhole solutions
with non-Abelian hair.

The wormhole solutions with an even number of nodes
possess always only a single throat.
In contrast, the wormhole solutions with an odd number of nodes
possess a double-throat with an equator inbetween,
when their size is sufficiently small.
In that case they also aquire a non-vanishing
surface gravity at their double-throat.

Since the equations depend only on the ratio $ \alpha$ of the
gravitational coupling $\bar \alpha$ and the size parameter $\eta_0^2$
of the throat, an increase in $ \alpha$ can be interpreted as
an increase of the gravitational coupling or a decrease of the
throat size. The limit $ \alpha \to \infty$ may thus be viewed
as the limit of vanishing throat size.

We have noted, that this limit is different for non-Abelian wormhole
solutions with an even or odd number of nodes.
The even solutions with $k=2n$ nodes converge smoothly towards
the Bartnik-McKinnon solutions with $n$ nodes
in the limit, while the spacetime splits up into two disjoint 
asymptotically flat parts.

The odd solutions with $k=2n+1$ nodes also converge towards
the Bartnik-McKinnon solutions with $n$ nodes,
however  this convergence is not smooth at the throat.
Instead a curvature singularity develops 
at the throat in this limit.
The convergence of the wormhole solutions to the Bartnik-McKinnon
solutions is also seen in the mass of the solutions.

The Bartnik-McKinnon solutions themselves are known to converge
towards a limiting solution for large node number $n \to \infty$.
This limiting solution consists of the extremal Reissner-Nordstr\"om solution
with unit charge in the outer region and a genuine non-Abelian solution
in the inner region, where both regions are joined at the extremal horizon.

Here we find a related behavior for the wormhole solutions
in the limit of large node number $k \to \infty$.
For $ \alpha < 1$ the non-Abelian solutions converge to the
corresponding embedded Abelian wormhole.
For $ \alpha > 1$, on the other hand, no embedded Abelian
wormhole solutions exist. Here - analogous to the Bartnik-McKinnon case -
the non-Abelian solutions converge towards
the extremal Reissner-Nordstr\"om solution
with unit charge in the outer region and a genuine non-Abelian solution
in the inner region, 
where again both regions are joined at the extremal horizon.

The analogy is even closer for the limiting behaviour of
the hairy black holes.
Here - in the region outside their horizon -
the hairy black holes converge towards the non-extremal
Abelian Reissner-Nordstr\"om
solution as long as their horizon is sufficiently large.
For horizon sizes smaller than the extremal 
Reissner-Nordstr\"om horizon, however, they converge towards
the extremal Reissner-Nordstr\"om solution
with unit charge in the outer region and a genuine non-Abelian solution
in the inner region, 
with both regions joined at the extremal horizon.

The wormhole solutions studied here possess the
reflection symmetry $\eta \to - \eta$. 
It will be interesting to consider 
also the general non-symmetric case.
Concerning their stability we note that
the Ellis wormhole is unstable
\cite{Shinkai:2002gv,Gonzalez:2008wd,Gonzalez:2008xk}.
Likewise, the Bartnik-McKinnon solutions 
and their associated hairy black holes are unstable
\cite{Straumann:1989tf,Straumann:1990as,Zhou:1991nu,Volkov:1994dq,Lavrelashvili:1994rp,Volkov:1995np}.
We therefore conjecture that the
hairy wormhole solutions will inherit these instabilties
and thus be unstable as well.

Regular Einstein-Yang-Mills solutions need not be spherically symmetric.
Indeed, static solutions are known, which possess only axial symmetry
\cite{Kleihaus:1996vi,Kleihaus:1997mn}.
These solutions give rise to families of static black holes,
whose horizon is only axially symmetric 
\cite{Kleihaus:1997ic,Kleihaus:1997ws},
showing that Israel's theorem 
does not generalize in the 
presence of non-Abelian fields.
We plan to construct the analogous set of hairy wormhole solutions,
which should be static and possess a throat that is only axially 
symmetric.
Presumably, also non-Abelian wormholes with throats
possessing only discrete symmetries may be found.

Einstein-Yang-Mills black holes may also rotate \cite{Kleihaus:2000kg}.
It would be interesting to construct 
the corresponding rotating Einstein-Yang-Mills wormholes.
Here as a first step rotating phantom field wormholes should be obtained.
These could then be employed to find the hairy solutions in the probe limit
and subsequently generalize them to include backreaction.

\subsection*{Acknowledgment}

We gratefully acknowledge discussions with E.~Radu and support 
by the DFG Research Training Group 1620 ``Models of Gravity''.

\end{document}